\documentclass[twocolumn,floats,floatfix,prd,superscriptaddress,nofootinbib,longbibliography]{revtex4-2}

\usepackage{graphicx,epsfig}
\usepackage{amssymb,amsmath,amsthm,amsfonts}
\usepackage{bm}
\usepackage[inline]{enumitem}
\usepackage{tensor}
\usepackage[linktocpage]{hyperref}
\usepackage[caption=false]{subfig}
\usepackage[usenames,dvipsnames]{xcolor}
\usepackage{url}
\usepackage[inline]{enumitem}
\usepackage{xspace}
\usepackage{comment}
\usepackage{cancel}
\usepackage{multirow}

\def\A{\mathbf{\mathcal{A}}}

\def\E{\mathcal{E}}
\def\F{\mathcal{F}}

\def\gB{\mathfrak{B}}
\def\gE{\mathfrak{E}}

\def\R{\mathcal{R}}

\def\Rm{\mathcal{R}^-}

\def\E{\mathcal{E}}

\def\bfR{\mathbf{R}}
\def\bfF{\mathbf{F}}
\def\bfg{\mathbf{g}}

\def\I{i\,}

\def\newacronym#1#2#3{\gdef#1{\gdef#1{#2\xspace}#3 (#2)\xspace}}
\newacronym{\bh}{BH}{black hole}
\newacronym{\bbh}{BBH}{black hole binary}
\newacronym{\gr}{GR}{General Relativity}
\newacronym{\bssn}{BSSN}{Baumgarte-Shapiro-Shibata-Nakamura}
\newacronym{\ts}{TS}{topological star}
\newacronym{\qnm}{QNM}{quasinormal mode}
\newacronym{\emd}{EMD}{Einstein-Maxwell-dilaton}
\newacronym{\nr}{NR}{Numerical Relativity}

\def\Julia{\textsc{Julia}\,}
\def\SCIML{\textsc{SciML}\,}
\def\DiffEq{\textsc{DifferentialEquations}\,}

\newcommand{\sapienza}{Dipartimento di Fisica, Sapienza Università 
	di Roma, Piazzale Aldo Moro 5, 00185, Roma, Italy}
\newcommand{\infn}{INFN, Sezione di Roma, Piazzale Aldo Moro 2, 00185, Roma, Italy}

\begin{document}

\title{Spectroscopy of magnetized black holes and topological stars
}

\author{Alexandru Dima}
\email{alexandru.dima@uniroma1.it}
\affiliation{\sapienza}
\affiliation{\infn}

\author{Marco Melis}
\email{marco.melis@uniroma1.it}
\affiliation{\sapienza}
\affiliation{\infn}

\author{Paolo Pani}
\email{paolo.pani@uniroma1.it}
\affiliation{\sapienza}
\affiliation{\infn}

\begin{abstract}
    We study the linear response of four dimensional magnetized black holes and regular topological stars arising from dimensional compactification of Einstein-Maxwell theory in five dimensions. We consider both radial and nonradial perturbations and study the stability of these solutions, both in the frequency and in the time domain. Due to the presence of magnetic fluxes in the background, axial (i.e., odd-parity) gravitational perturbations are coupled to polar (i.e., even-parity) electromagnetic perturbations (Type-I sector) whereas polar gravitational and scalar perturbations are coupled to axial electromagnetic  ones (Type-II sector). We provide a comprehensive analytical and numerical study of the radial perturbations and of the Type I sector, finding no evidence of linear instabilities (besides the already known Gregory-Laflamme instability of black strings occurring only in a certain range of the parameters), even despite the fact that the effective potential for radial perturbations of topological stars is negative and divergent near the inner boundary.
    Ultracompact topological stars exhibit long-lived trapped modes that give rise to echoes in the time-domain response. Their prompt response is very similar to that of the corresponding black hole with comparable charge-to-mass ratio. This provides a concrete realization of ultracompact objects arising from a well-defined theory.
    The numerical analysis of the Type-II sector will appear in a companion paper.  
\end{abstract}

\maketitle

\tableofcontents

\begin{figure*}[th]
  \centering
\includegraphics[width=0.495\textwidth]{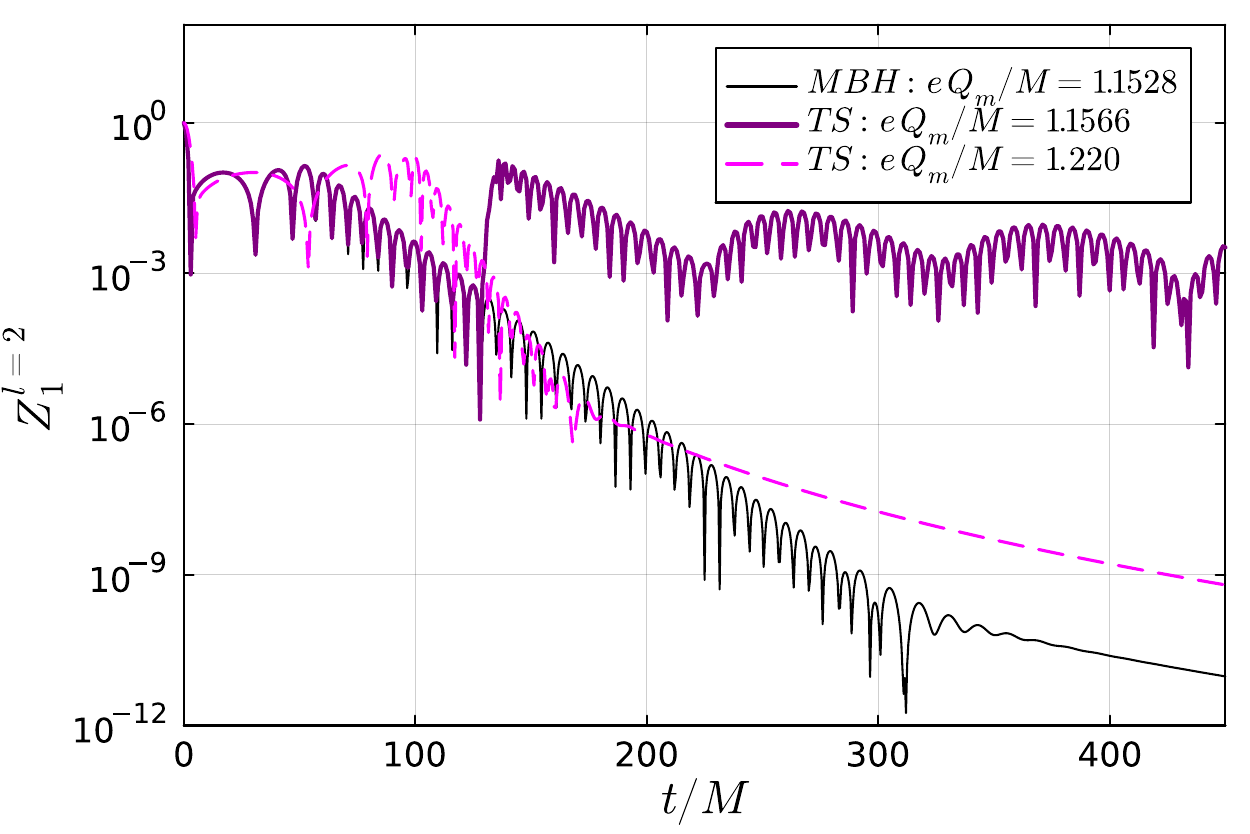}
\includegraphics[width=0.495\textwidth]{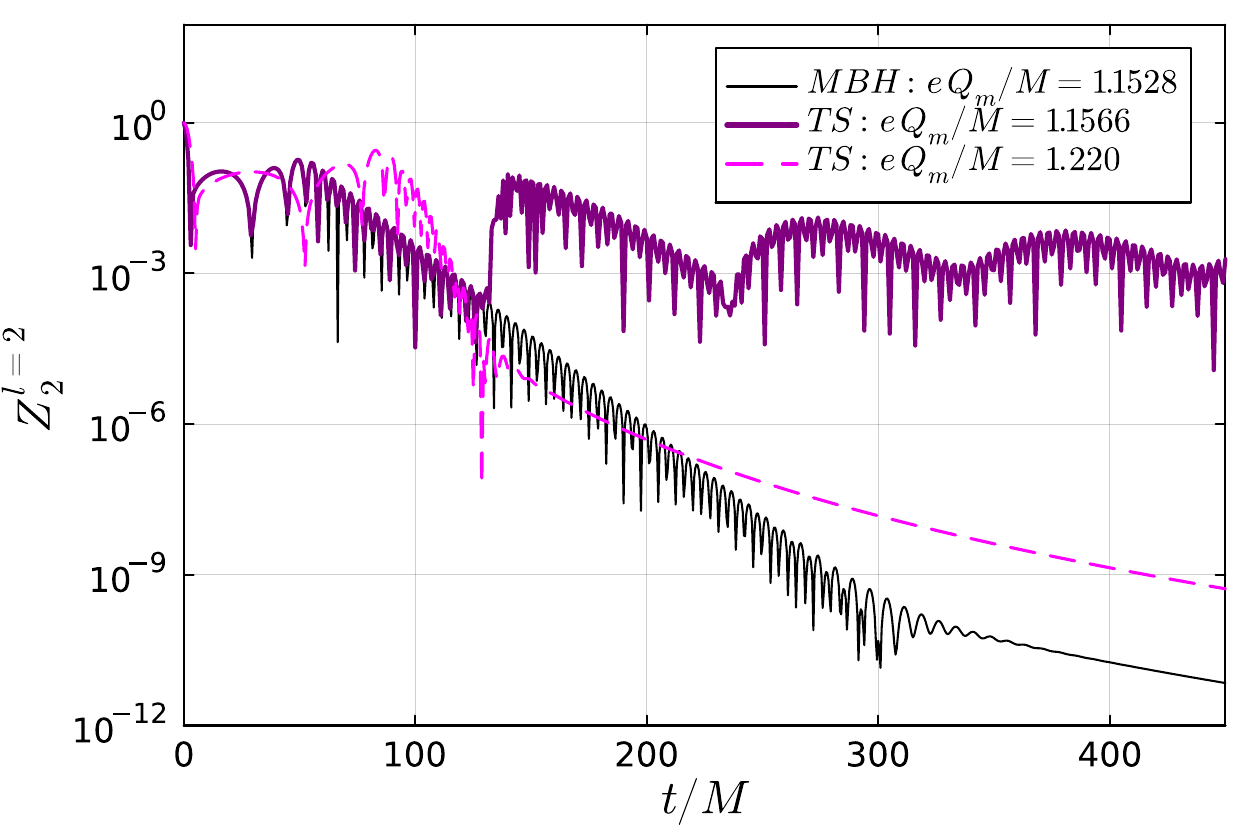}
  \caption{Comparison between the linear response of a nearly-extremal magnetized BH and a second-kind TS with same mass and similar charge-to-mass ratio to $l=2$ Type-I perturbations (even-parity EM, and odd-parity gravitational). The left and right panel refer to gravitational-driven and EM-driven perturbations, respectively. 
  For comparison we show also the case of a first-kind TS, which displays a different prompt ringdown and no long-lived modes.
  } \label{fig:Z12_RingdownVSEchoes}
\end{figure*}

\section{Introduction}
The fuzzball program of string theory aims at describing the classical black hole~(BH) horizon as a coarse-grained description of a superposition of regular quantum states~\cite{Mathur:2009hf,Bena:2022rna,Bena:2022ldq}. The horizon-scale structure is provided by ``microstate geometries'': solitons with the same mass and charges as a BH, but where the horizon is replaced by a smooth horizonless cap~\cite{Bena:2006kb,Bena:2016ypk,Bena:2017xbt,Bah:2021owp,Bah:2022yji}. These solutions hinge on two distinctive features of string theory: the presence of non-perturbative entities known as D-branes, whose mass diminishes with increasing gravitational strength, and the introduction of numerous new degrees of freedom (possibly accounting for the BH entropy), which boost quantum tunneling and prevent the formation of horizons~\cite{Kraus:2015zda,Bena:2015dpt}.

A key aspect of this program is that it is intrinsically higher dimensional and relies on nontrivial topologies to prevent the horizon-scale structure from collapsing. The microstates characterizing horizon-scale structure are smooth, topologically nontrivial geometries in ten dimensions. However, when reduced to four dimensions,
they appear to have curvature singularities~\cite{Balasubramanian:2006gi,Bena:2022fzf}. 
Dimensional reduction also gives rise to a number of Kaluza-Klein fields nonminimally coupled to gauge fields.
Furthermore, regardless of the details of horizon-scale structure, the large Hilbert space describing the states that give rise to the huge BH entropy will contain coherent states. These will resemble classical solutions in low-energy theories coupled with gravity. 
Thus, from a four-dimensional perspective, one is left with General Relativity coupled to various forms of matter, including gauge fields and scalars, and potential singularities which are anyway well behaved from the 5D perspective.

Einstein-Maxwell theory in five dimensions allows for magnetized black strings and regular solitons known as topological stars~(TSs)~\cite{Bah:2020ogh}. The scope of this paper is to study the linearized dynamics and stability of these objects as a toy model for more complicated and realistic microstate geometries.
These solutions are particularly interesting because they contain several ingredients of more complicated microstate geometries while keeping a certain degree of symmetry and hence being more tractable. In particular, while being regular in five dimensions, from a four dimensional perspective they contain an extra scalar field that diverges at the boundary of the TS solution, where also the metric becomes singular. This implies that extra care should be put in investigating the boundary conditions (BC) in the four-dimensional theory.
Furthermore, the spherically symmetric solution has a magnetic field
which mixes sectors with different parity. In general, scalar, electromagnetic~(EM), and gravitational perturbations are coupled to each other in a nontrivial way, as generically expected from classical solutions of a low-energy effective theory.

Due to their nontrivial structure, a natural question concerns the stability of these solutions.
Linear perturbations of magnetized BHs in this theory were partially studied in~\cite{Guo:2022rms,Guo:2023vmc}.
Due to the presence of magnetic fluxes in the background, polar (i.e., even-parity) gravitational perturbations are coupled to axial (i.e., odd-parity,) EM perturbations and viceversa.
We shall refer to the sector containing odd-parity (resp. even-parity) gravitational perturbations as Type-I (resp.~Type-II). In~\cite{Guo:2023vmc}, the quasinormal modes~(QNMs) of magnetized BHs in this theory were obtained for the Type-I sector, which is  easier than the Type-II sector since it contains less dynamical degrees of freedom.
For the case of TSs, only the linear dynamics of a test scalar field in the frequency domain has been studied~\cite{Heidmann:2023ojf,Bianchi:2023sfs}, finding different families of modes depending on parameters of the TS.
In particular, in some regions of the parameter space, TSs can develop a pair of stable and unstable photon spheres which can support long-lived modes~\cite{Cardoso:2014sna} and can give rise to echoes~\cite{Cardoso:2016oxy,Cardoso:2016rao,Cardoso:2017cqb,Abedi:2020ujo} in the ringdown signal at late times.
From this perspective, TSs provide a concrete model for ultracompact objects~\cite{Cardoso:2019rvt} arising from a well-defined theory and are therefore an ideal testbed to investigate the phenomenology of these objects. Furthermore, since TSs are singular from a four-dimensional perspective, they have quite unique properties relative to other phenomenological models of ultracompact objects, and indeed they are consistent solutions only when considered as arising from a higher-dimensional theory upon dimensional reduction.

Here, we greatly extend this program by studying the complete linearized dynamics (in which scalar, EM, and gravitational perturbations are coupled to each other) both in the frequency and in the time domain. This will also allow us to discuss the linear stability of magnetized BHs and TSs.
While we will derive the equations for all kinds of perturbations, in this work we numerical solve for the dynamics of radial perturbations and of nonradial perturbations in the Type-I sector. Nonradial Type-II perturbations will be studied in a companion paper~\cite{companion}.
Overall, in the radial Type-II and nonradial Type-I sectors we found no evidence for linear instabilities\footnote{Beside a known radial instability~\cite{Miyamoto:2006nd,Stotyn:2011tv,Bah:2021irr} associated to the Gregory-Laflamme instability of black strings~\cite{Gregory:1993vy}, which occurs only in a certain range of the parameters, see below for further details.}, even despite the fact that the effective potential for radial perturbations of TSs is negative and divergent near the inner boundary. For both BHs and TSs, the QNMs computed in the frequency domain are in perfect agreement with the object's response to small perturbations computed in the time domain.
As found in Ref.~\cite{Heidmann:2023ojf} for test scalar perturbations, we find that TSs without a stable photon sphere have (scalar, EM, and gravitational) QNMs similar to those of BHs, which indeed resemble the gravitational $w$-modes of compact stars in General Relativity~\cite{Kokkotas:1999bd}.
TSs with a pair of stable and unstable photon spheres have long-lived QNMs that dominate the object response at late time. As expected for ultracompact objects~\cite{Cardoso:2019rvt}, the response in the time domain is initially very similar to that of a BH with similar charge-to-mass ratio, while the late time response is governed by echoes.
An example of this behavior is anticipated in Fig.~\ref{fig:Z12_RingdownVSEchoes}, which will be discussed in detail in the rest of the paper.
The same effect was observed for various classes of ultracompact objects (see, e.g.,~\cite{Cardoso:2016oxy,Cardoso:2016rao,Abedi:2016hgu,Mark:2017dnq,Nakano:2017fvh,Bueno:2017hyj,Wang:2018gin,Raposo:2018rjn,Pani:2018flj,Konoplya:2018yrp,Cardoso:2019apo,Maggio:2019zyv,Conklin:2019fcs,Wang:2019rcf,Oshita:2019sat,Dey:2020lhq,Maggio:2020jml,Chakraborty:2022zlq}), including BH microstates~\cite{Ikeda:2021uvc} (see also \cite{doi:10.1098/rspa.1991.0104, 10.1093/mnras/268.4.1015, PhysRevD.62.107504} for earlier studies on trapped modes in ultracompact incompressible stars). However, in many of the previous studies the background solution was either phenomenological or pathological, while in~\cite{Ikeda:2021uvc} only test scalar perturbations were studied, due to the complexity of the theory.
To the best of our knowledge this is the first example of clean echoes appearing in the gravitational waves emitted by a consistent and stable solution to a well-defined theory.

The rest of the paper is organized as follows.
Section~\ref{sec:setup} presents the setup and the various sectors of the linearized field equations, in many cases providing them in Schr\"{o}dinger-like form with an analytical effective potential.
Section~\ref{sec:results} presents our numerical results for the QNMs and linear response in time of magnetized BHs and TSs.
We conclude in Sec.~\ref{sec:conclusion} and provide some technical details of the computations in the appendices.

{\bf Note added:}
While this work was nearly completion, we were informed that another group was working independently on the same problem~\cite{Bena:2024hoh}. Although there is significant overlap, our analysis and that of~\cite{Bena:2024hoh} also focus on different aspects and numerical methods and are therefore complementary to each other. We have compared several numerical results with those of~\cite{Bena:2024hoh}, finding excellent agreement, especially for long-lived modes.

\section{Setup and master equations} \label{sec:setup}
\subsection{Five-dimensional theory, field equations, and background solutions}
We consider Einstein-Maxwell theory in 5D,
\begin{equation}
    S_5 = \int d^5 x \sqrt{-\mathbf{g}} \left(\frac{1}{2\kappa_5^2}\bfR - \frac{1}{4}\bfF_{AB}\bfF^{AB} \right)\,,
\end{equation}
yielding the covariant equations:
\begin{align}
   & \bfR_{AB}-\frac{1}{2}\bfg_{AB} \bfR + \kappa_5^2 \left( \bfF_{AC}\bfF^{C}{}_{B} + \frac{1}{4}\bfg_{AB}\bfF_{CD}\bfF^{CD} \right) = 0
   \\
   & \mathbf{\nabla}^B\bfF_{AB} = 0
\end{align}
This theory admits a regular solution known as TS~\cite{Bah:2020ogh}
\begin{align} ~\label{eq:BH/TS_5D_metric}
ds^2 & =-f_S dt^2+f_Bdy^2+\frac{1}{h}dr^2+r^2d\Omega_2^2
\\
F & = P \sin\theta \,d\theta \wedge d\phi
\end{align}
where
\begin{align}
&f_S=1-\frac{r_S}{r}\,, \hspace{1em}f_B=1-\frac{r_B}{r}\,,
\nonumber\\
& h=f_B f_S\,,\hspace{1em} P = \pm \frac{1}{\kappa_5}\sqrt{\frac{3r_Sr_B}{2}}\,.
\end{align}
with $r_B > r_S$.
The solution is everywhere regular\footnote{The parameters $r_S$, $r_B$, and $R_y$ are algebraically constrained by the orbifold condition, which also implies $2r_B\leq n R_y$, where $n$ is an integer~\cite{Bah:2020ogh}. For $n=1$ the solution is smooth, whereas for $n>1$ it contains a conical defect, which can be anyway smoothly resolved if the solution is embedded in a string theory~\cite{Bah:2020ogh}. The phenomenologically interesting range in which the extra dimension is small ($R_y\ll r_B$) imposes $n\gg1$. Alternatively, one can consider $n=1$ and stack $N\gg1$ topological stars, in which case the solution has a size $r_B\sim N^{1/4} R_y \gg R_y$ and has no conical defect~\cite{Bah:2021rki}.}

and asymptotes to four dimensional Minkowski times a circle, parametrized by the coordinate $y$ with period $2\pi R_y$.
The case $r_B\leq r_S$ corresponds to a magnetized black string with event horizon located at $r=r_S$. In the following we shall study the linear perturbations of both solutions.

\subsection{Four-dimensional compactification}
~\label{sec:KKcompactification}
To study the linear perturbations of magnetized BHs and TSs, we perform a four-dimensional compactification, 
introducing a scalar field $\Phi$ and a gauge field ${\cal A}_\mu$ for the gravity sector, and a scalar field $\Xi$ for the EM sector:
\begin{align}
    ds^2_5&= e^{-\frac{\sqrt{3}}{3} \Phi}ds_4^2 + e^{2\frac{\sqrt{3}}{3} \Phi}(dy+\A_\mu dx^\mu)^2\,,\\
    \mathbf{F}_{AB}dx^Adx^B&=F_{\mu\nu}dx^\mu dx^\nu + \left( \partial_\mu\Xi dx^\mu \right) \wedge \left( dy + A_\mu dx^\mu \right) \,,
\end{align}
where henceforth we define the 4D field strengths ${\cal F}_{\mu\nu}=\partial_\mu {\cal A}_\nu-\partial_\nu {\cal A}_\mu$ and ${F}_{\mu\nu}=\partial_\mu {A}_\nu-\partial_\nu {A}_\mu$.
We assume that all variables are independent of the extra dimension $y$. While this is certainly true for the background solution, the translation symmetry along $y$ of the latter implies that perturbations can be decomposed with a $e^{iky}$ dependence, where $k= p/R_y$ is the quantized momentum along $y$ and $p=0,1,2,..$.
Phenomenologically, one is interested in the case in which the extra dimension is small and the solution is macroscopic. This requires $R_y\ll r_S$, so perturbations with $p\neq0$ are hardly excited in classical processes if the object is macroscopic. Henceforth, we will assume $p=0$, so there is no $y$ dependence in the dynamical variables\footnote{Later on we will briefly discuss radial perturbations with nonvanishing Kaluza-Klein momentum, which are relevant for the Gregory-Laflamme instability of a black string~\cite{Gregory:1993vy}.}.

In this setup, the corresponding 4D action describes an Einstein-Maxwell-Dilaton~(EMD) theory with two scalars and two gauge fields: 
\begin{align}~\label{eq:EMS4DAction}
     \mathcal{S} 
     & = \int dx^4 \sqrt{-g}\left[
     \frac{1}{2\kappa_4^2}\left( R - \frac{1}{2} \partial_\mu\Phi\partial^\mu\Phi -\frac{1}{4}e^{\sqrt{3}\Phi}\F_{\mu\nu}\F^{\mu\nu}\right)\right. \nonumber\\
     &+\left.
     \frac{1}{e^2}\left( -\frac{1}{4}e^{\frac{\sqrt{3}}{3}\Phi} F_{\mu\nu}F^{\mu\nu} -\frac{1}{2}e^{-\frac{2\sqrt{3}}{3}\Phi}(\partial_\mu \Xi)^2\right)
     \right]
     \,,
\end{align}
giving the 4D field equations:
\begin{widetext}
\begin{align}
   & G_{\mu\nu}  
   + \left[ \frac{1}{2}e^{\sqrt{3}\Phi}\left( \F_{\mu\rho}\F^{\rho}{}_{\nu} + \frac{1}{4}g_{\mu\nu}\F_{\rho\sigma}\F^{\rho\sigma} \right)
    - \frac{1}{2}\left( \partial_\mu\Phi\partial_\nu\Phi -\frac{1}{2} g_{\mu\nu}\partial_\rho\Phi\partial^\rho\Phi\right) \right]
   \nonumber\\
   &
 + \frac{\kappa_4^2}{e^2} \left[ e^{\frac{\Phi}{\sqrt{3}}}\left( F_{\mu\rho}F^{\rho}{}_{\nu} + \frac{1}{4}g_{\mu\nu}F_{\rho\sigma}F^{\rho\sigma} \right)
 - e^{-\frac{2\Phi}{\sqrt{3}}}\left(\partial_\mu\Xi\partial_\nu\Xi-\frac{1}{2}g_{\mu\nu}\partial^\rho\Xi\partial_\rho\Xi\right)\right] = 0 
    \label{eq:EMSEinsteinEq}
   \,,\\
   & \nabla^\rho \left(e^{\sqrt{3}\Phi}\F_{\mu\rho}\right)  = 0
    \label{eq:EMScurlyFmunu}
    \,,\\
& \Box\Phi  
- \frac{\sqrt{3}}{4}e^{\sqrt{3}\Phi}\F_{\mu\nu}\F^{\mu\nu} 
+\frac{\kappa_4^2}{e^2}\left[ \frac{2\sqrt{3}}{3}e^{-\frac{2\Phi}{\sqrt{3}}}\left(\partial_\mu\Xi\right)^2 
- \frac{\sqrt{3}}{6}e^{\frac{\Phi}{\sqrt{3}}}F_{\mu\nu}F^{\mu\nu} \right]
= 0
   ~\label{eq:EMSPhi}
    \,,\\
    & \nabla^\rho \left(e^{\frac{\sqrt{3}}{3}\Phi} F_{\mu\rho}\right) = 0
       ~\label{eq:EMSFmunu}
    \,,\\
   & \nabla^\rho \left(e^{-\frac{2\sqrt{3}}{3}\Phi}\partial_\rho\Xi\right) =0
   ~\label{eq:EMSxi}
    \,.
\end{align}
\end{widetext}
We introduced here the couplings $\kappa_4^2:=\kappa_5^2/(2\pi R_y)$ and $e^2:=1/(2\pi R_y)$, where $R_y$ is the radius of the compact extra dimension\footnote{We will keep these coupling constants explicit but, when presenting numerical results, we will use units such that $\kappa_4=\sqrt{8\pi}$ and $e=\sqrt{4\pi}$.}.
The dynamics of the 4D theory, which includes gravity, two gauge fields ${\cal A}_\mu$ and $A_\mu$, and two scalar fields $\Phi$ and $\Xi$, is fully equivalent to the 5D Einstein-Maxwell theory for modes with no dependence along the fifth direction.

The background 4D line element reads
\begin{align}
ds_4^2 & =-f_S f_B^{1/2}dt^2+\frac{1}{f_S f_B^{1/2}}dr^2+r^2 f_B^{1/2}d\Omega_2^2 \,, ~\label{eq:BH/TS_bg_metric}
\\
\Phi & = \frac{\sqrt{3}}{2} \log f_B \,,
\\
F & = \pm e Q_m\sin\theta \,d\theta \wedge d\phi
\nonumber\\
& = \pm \frac{e}{\kappa_4}\sqrt{\frac{3}{2}r_Br_S}\sin\theta \,d\theta \wedge d\phi
\\
\F & =0=\Xi ~\label{eq:BH/TS_bg0}
\,.
\end{align}

Note that both the metric and the dilaton diverge at $r=r_B$, but the TS solution is regular in the 5D uplift.
The ADM mass and magnetic charge of the background are, respectively,
\begin{equation}
    M = \frac{2\pi}{\kappa_4^2}(2r_S+r_B)
    \,,
    \hspace{1em}
    Q_m = \frac{1}{\kappa_4}\sqrt{\frac{3}{2}r_Sr_B}
    \,.
\end{equation}

\begin{figure}[th]
  \centering
  \includegraphics[width=0.45\textwidth]{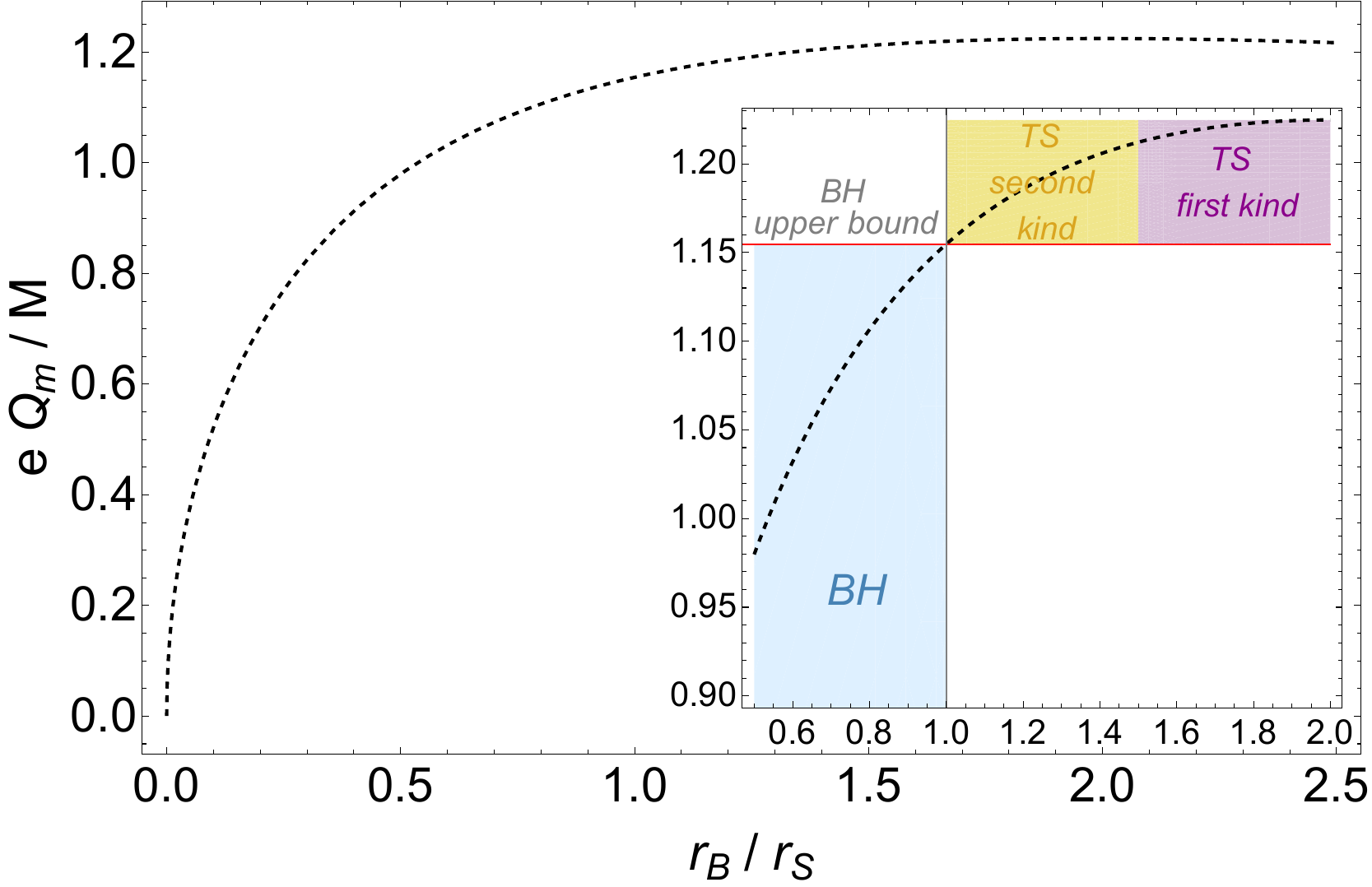}
  \caption{Parameter space of static magnetized BHs and topological stars.
  The inset shows the parameter space where solutions do not suffer from known instabilities.
  }\label{fig:paramspace}
\end{figure}

The parameter space of static magnetized BHs and TSs in this theory is depicted in Fig.~\ref{fig:paramspace}. This is obtained by inverting the above relations to get $Q_m/M$ as a function of $r_B/r_S$.

Magnetized BHs require $r_B/r_S<1$ which implies
$\frac{eQ_m}{M} \leq \frac{e\kappa_4}{2 \pi \sqrt{6}}\approx 1.1547$. When $0<r_B/r_S<1/2$, these solutions are linearly unstable against the Gregory-Laflamme mechanism~\cite{Miyamoto:2006nd,Gregory:1993vy}.
For TSs ($r_B/r_S>1$) we have $\frac{e\kappa_4}{2 \pi \sqrt{6}} \leq \frac{eQ_m}{M} \leq \frac{\sqrt{3}e\kappa_4}{8 \pi}$. Solutions with $r_B/r_S>2$ are also unstable~\cite{Stotyn:2011tv,Bah:2021irr}, as can be obtained from the aforementioned Gregory-Laflamme instability of magnetized BHs and performing a double Wick rotation 
$(t, y, r_S , r_B) \to (iy, it, r_B , r_S)$ 
which maps BHs to TSs~\cite{Bah:2021irr}. We will confirm this result numerically in Sec.~\ref{sec:radial}.

Note that both magnetized BHs and TSs can have $eQ_m/M>1$, at variance with the four-dimensional Reissner-Nordstr\"{o}m BH, which is not a solution to this theory.

While the magnetized BH is characterized by a single unstable photon sphere at $r_{\rm ph} = \frac32 r_S$, the TS may show a pair of stable and unstable photon spheres, depending on the parameter space~\cite{Heidmann:2023ojf,Bianchi:2023sfs,Lim:2021ejg}. We can classify TSs in two families:
\begin{align}
    \textit{TS first kind}, \hspace{0.3cm} \frac32 < \frac{r_B}{r_S} \leq 2 : \hspace{0.5cm} &r_{\rm ph}^{(1)} = r_B \,, \\
    \textit{TS second kind}, \hspace{0.3cm} 1 \leq \frac{r_B}{r_S} \leq \frac32 : \hspace{0.5cm}  &r_{\rm ph}^{(1)} = \frac32 r_S \,, \hspace{0.3cm} r_{\rm ph}^{(2)} = r_B \,.
\end{align}
TSs of the second kind have a \emph{stable} photon sphere at $r_{\rm ph}^{(2)} = r_B$ and an unstable one at $r_{\rm ph}^{(1)} = \frac32 r_S$, just like the magnetized BH solution.

\subsection{Linear dynamics}

In Appendix~\ref{app:RW} we perform a Regge-Wheeler-Zerilli~\cite{Regge:1957td,Zerilli:1970se} spherical-harmonic decomposition of the metric, EM, and scalar perturbations of the background.
The magnetic field of the background breaks parity and enforces a coupling between even-parity gravito-scalar perturbations and odd-parity EM perturbations, and viceversa~\footnote{See~\cite{Pereniguez:2023wxf} for a generalized method based on the electric-magnetic duality that allows to decouple perturbations with different parities.} .
We discuss different independent sectors below.

Since ${\cal F}=0=\Xi$ on the background, the perturbations of these fields decouple from the others. Through a field redefinition, at the linear level Eqs.~\eqref{eq:EMScurlyFmunu} and \eqref{eq:EMSxi} can be written as those for a test Maxwell and massless scalar field, respectively, propagating on the fixed background. 
Since test scalar perturbations of TSs and BHs have been studied in~\cite{Heidmann:2023ojf,Bianchi:2023sfs} and Maxwell perturbations of a magnetized BH have been studied in~\cite{Guo:2023vmc}, here we do not discuss these further, focusing instead to the coupled system of equations. For completeness, the field equations for the decoupled vector and scalar fields are given in Appendix~\ref{app:decoupled}.

We will present the main equations in the frequency domain, assuming a $\sim e^{-i\omega t}$ time dependence for each variable. These are the equations that will be relevant to compute the QNMs as a one-dimensional eigenvalue problem. When needed, in the end we will present also the time-domain version of the relevant equations that will be integrated with a $1+1$ evolution code.

\subsubsection{Type-I perturbations} \label{sec:eqTypeI}
The Type-I sector couples odd-parity metric components with even-parity EM components and is decoupled from the scalar perturbations. To derive the evolution equations we adopt and generalize the approach used in~\cite{Nomura:2020tpc}.
For $l\geq 2$ perturbations, we obtain a system of two equations 
\begin{widetext}
\begin{align}
&~\label{eq:Rm_B&Hcoordbis}
{\cal D}[\Rm(r)] + \left( 2 f_S^2f_B' + 3 f_Bf_Sf_S' \right) \partial_r \Rm(r) \\ \nonumber
&
- \left(
\frac{2f_Bf_S^2}{r^2} 
+ \frac{\left( \Lambda - 2 \right)f_S}{r^2} 
+ \frac{2f_S^2f_B' + f_B f_S f_S'}{r} 
- 3f_Sf_B'f_S'
- f_Bf_S'^2
\right) \Rm(r)
- \frac{2\kappa_4^2 Q_m}{e r^3} \E(r) = 0
\\
&~\label{eq:E_B&Hcoordbis}
{\cal D}[\E(r)] + \left( 2f_S^2f_B' + f_Bf_Sf_S' \right) \partial_r \E(r) \\ \nonumber
&
- \left(
 \frac{2\kappa_4^2Q^2_mf_S}{r^4} 
+ \frac{\Lambda f_S}{r^2} 
+ \frac{2f_S^2f_B'}{r}
- f_Sf_B'f_S'
\right)\E(r) 
- \frac{eQ_m\left(\Lambda-2\right)f_S^2}{r^3}\Rm(r) = 0
\end{align}
\end{widetext}
where we defined the second-order differential operator
\begin{equation}
    {\cal D}=\left(f_Bf_S^2\right) \partial_r^2 + \omega^2\,,\label{eq:diffop} 
\end{equation}
and
$\Lambda=l(l+1)$. 
To obtain the above system we combine the $(r, \phi)$ and $(\theta, \phi)$ components of the perturbed Einstein equation~\eqref{eq:EMSEinsteinEq} together with the radial component of the perturbed Maxwell equation~\eqref{eq:EMSFmunu} to get a second-order equation for $h_1$ sourced only by $f_{01}^+$. Analogously, we can derive a second-order equation for $f_{01}^+$ from the $t$ and $r$ components of the Maxwell equations, combined with the Maxwell constraint $f_{01}^+-\I\omega f_{12}^+ - \partial_r f_{02}^+=0$.
The relation between metric and EM perturbations defined in Appendix~\ref{app:RW} and the auxiliary variables $\Rm$ and $\E$ is given in terms of
\begin{align}
& h_1(r) = -\I \omega r \sqrt{f_B} \Rm(r)\,,
\\
& f_{01}^+(r) =\frac{\Lambda}{r^2}\E(r)\,.
\end{align}

Interestingly, the above equations can be decoupled. One can first introduce a generalized  tortoise coordinate defined by 
\begin{align}
\label{eq:gentortoise}
        d\rho = \sqrt{\frac{g_{rr}}{g_{tt}}}dr = \frac{dr}{f_B^{1/2}f_S}\,,
\end{align}
which can be integrated to obtain a closed-form expression\footnote{This expression is valid both when $r_B>r_S$ and when $r_S>r_B$. For the latter case (magnetized BHs) the integration constant should be fixed to ensure that $\rho(r)$ is a real function outside the horizon, $r>r_S$.} for $\rho(r)$.
Then, making a field redefinition
\begin{equation}
    \Tilde{\R}^-(r) = f_B^{3/4} f_S \Rm(r) \,, \quad \Tilde{\E}(r) = f_B^{3/4}\frac{{\kappa}_4}{e}\sqrt{\frac{2}{\Lambda-2}} \E(r) \,,
\end{equation}
we obtain 
\begin{align}
    \left( \frac{d^2}{d\rho^2} + \omega^2 \right)\begin{pmatrix}
        \Tilde{\R}^- \\
        \Tilde{\E}
    \end{pmatrix}
    = \textbf{B} \begin{pmatrix}
        \Tilde{\R}^- \\
        \Tilde{\E}
    \end{pmatrix} \,,
\end{align}
where 
\begin{align}
    \textbf{B} = \frac{f_S}{r^3}\left[ F(r)\begin{pmatrix}
        1 & 0 \\ 
        0 & 1
    \end{pmatrix} 
    +\begin{pmatrix}
        0 & P \\ 
        P & 2 r_B + 3 r_S
    \end{pmatrix}
    \right] \,,
\end{align}
where $F(r)=\Lambda r - \frac{3(13 r_B^2 r_S + 8r^2(r_B+2r_S)-r r_B(9r_B+28r_S))}{16 r (r-r_B)}$ and $P=\sqrt{3(\Lambda-2)r_B r_S}$.
The system above can be decoupled by performing a linear, $r-$independent transformation 
\begin{align}
    Z_1 &= \mathcal{L}_1 \Tilde{\R}^- + \mathcal{L}_2 \Tilde{\E} \\
    Z_2 &= \mathcal{L}_2 \Tilde{\R}^- - \mathcal{L}_1 \Tilde{\E}
\end{align}
with
\begin{align}
    \mathcal{L}_1 &= -(2 r_B + 3 r_S) - \sqrt{(2r_B + 3r_S)^2+12 \Lambda r_S r_B} \,, \\
    \mathcal{L}_2 &= 2 \sqrt{3 (\Lambda-2) r_S r_B} \,.
\end{align}
The decoupled system is 
\begin{align}
    \left( \frac{d^2}{d\rho^2} + \omega^2 \right) Z_i
    = V_{\rm eff}^i Z_i \hspace{1cm} i = 1,2 \label{eq:typeIdecoupled}
\end{align}
with the effective potentials
\begin{align}
    V_{\rm eff}^{1,2} &= \frac{r-r_S}{16 r^5 (r-r_B)} [16 r^3 \Lambda - r^2 (8r_B + 24 r_S + 16 \Lambda r_B) \nonumber\\
    &+r(11 r_B^2 + 60 r_B r_S) - 39 r_B^2 r_S  \nonumber\\
    &\mp 8 r (r-r_B)\sqrt{(2 r_B - 3 r_S)^2+ 12r_B r_S \Lambda}] \,
\end{align}

where the minus and the plus signs correspond to $V_{\rm eff}^{1}$ and $V_{\rm eff}^{2}$, respectively. Finally, for $l=1$, the metric perturbation $h_0$ can be eliminated via a gauge transformation and the $h_1$ perturbation is nondynamical. Indeed, by the same combination of equations used to derive Eqs.~\eqref{eq:Rm_B&Hcoordbis} and ~\eqref{eq:E_B&Hcoordbis}, in this case one can check that $h_1$ can be fixed as a function of $f_{01}^+$ via the relation 
\begin{align}
    h_1 = \frac{Q_m \kappa_4^2 f_B^{1/2}}{\I \omega e} f_{01}^+ \,.
\end{align}

Thus, for $l=1$ the Type-I sector reduces to a single master equation: 
\begin{align}
&~\label{eq:E_B&Hcoord_l1}
{\cal D}[\E(r)] + (2 f_S^2 f_B' + f_B f_S f_S')\partial_r \E(r) 
-V_{\rm eff}^{l=1} \E(r)  = 0\,,
\end{align}
where $f_{01}^+(t,r) =\frac{1}{r^2}\E(t,r)$
and the effective potential reads
\begin{equation}
V_{\rm eff}^{l=1}=f_S\left(  
 \frac{2\kappa_4^2Q^2_m}{r^4} 
+ \frac{2}{r^2} 
+ \frac{2f_Sf_B'}{r}
- f_B'f_S'\right)\,.
\end{equation}

\subsubsection{Type-II perturbations}

The Type-II sector is more involved because it couples even-parity metric components and the scalar perturbations with odd-parity EM components.
In Appendix~\ref{app:type_II} we list the relevant components of the perturbed Einstein equations, along with the scalar and axial gauge perturbation equations. The discussion and solution of these equations is deferred to a companion paper, which will address the analysis of Type-II perturbations in the general case for $l \geq 1$. Instead, in the next section we will focus on radial ($l=0$) perturbations.

\subsubsection{Radial perturbations} \label{sec:radial}
Radial perturbations belong to the Type-II sector but, because $l=0$ gravitational and EM perturbations are nondynamical, they are much easier to study.
In the radial case the metric perturbation $K$ in front of the two-sphere submanifold and the off-diagonal perturbation $H_1$ can be eliminated with a coordinate choice, so one is left only with the diagonal perturbations of the $(t,r)$ submanifold, namely $H_0$ and $H_2$. Furthermore, the only radial mode of the EM sector is even-parity, so it does not contribute to Type-II perturbations. Using Einstein's equations, one finds two constraints relating $H_0$ and $H_2$ to the dynamical scalar perturbation, which is governed by a single master equation: 

\begin{align}
   & {\cal D}[\varphi]
   + \left( f_S^2f_B'+f_Bf_Sf_S' \right) \partial_r\varphi
   - V_{\rm eff}^{l=0} \varphi
   =0\,,
 \label{eq:typeII_phic}
\end{align}
with the effective potential
\begin{align}
    V_{\rm eff}^{l=0}(r)&=\frac{f_S^2f_B'+f_Bf_Sf_S'}{r} 
   + \frac{Q_m^2\kappa_4^2 f_S}{3r^4} 
   + \frac{Q_m^2\kappa_4^2f_Sf_B'}{2r^3(4f_B+rf_B')}\nonumber\\
   & - \frac{6f_S^2f_B'^2(f_B+rf_B')}{(4f_B+rf_B')^2}
   + \frac{3 r f_S f_B'^2 f_S'}{4(4f_B+rf_B')}\,.
\end{align}
As discussed below, in the TS case the potential diverges at the boundary, $V_{\rm eff}^{l=0}(r\to r_B)\to-\infty$, is zero at some $r>r_B$, and vanishes at spatial infinity. In the magnetized BH case the potential vanishes also at $r=r_S$ and has the standard shape.

\begin{figure*}[th]
  \centering
\includegraphics[width=1\textwidth]{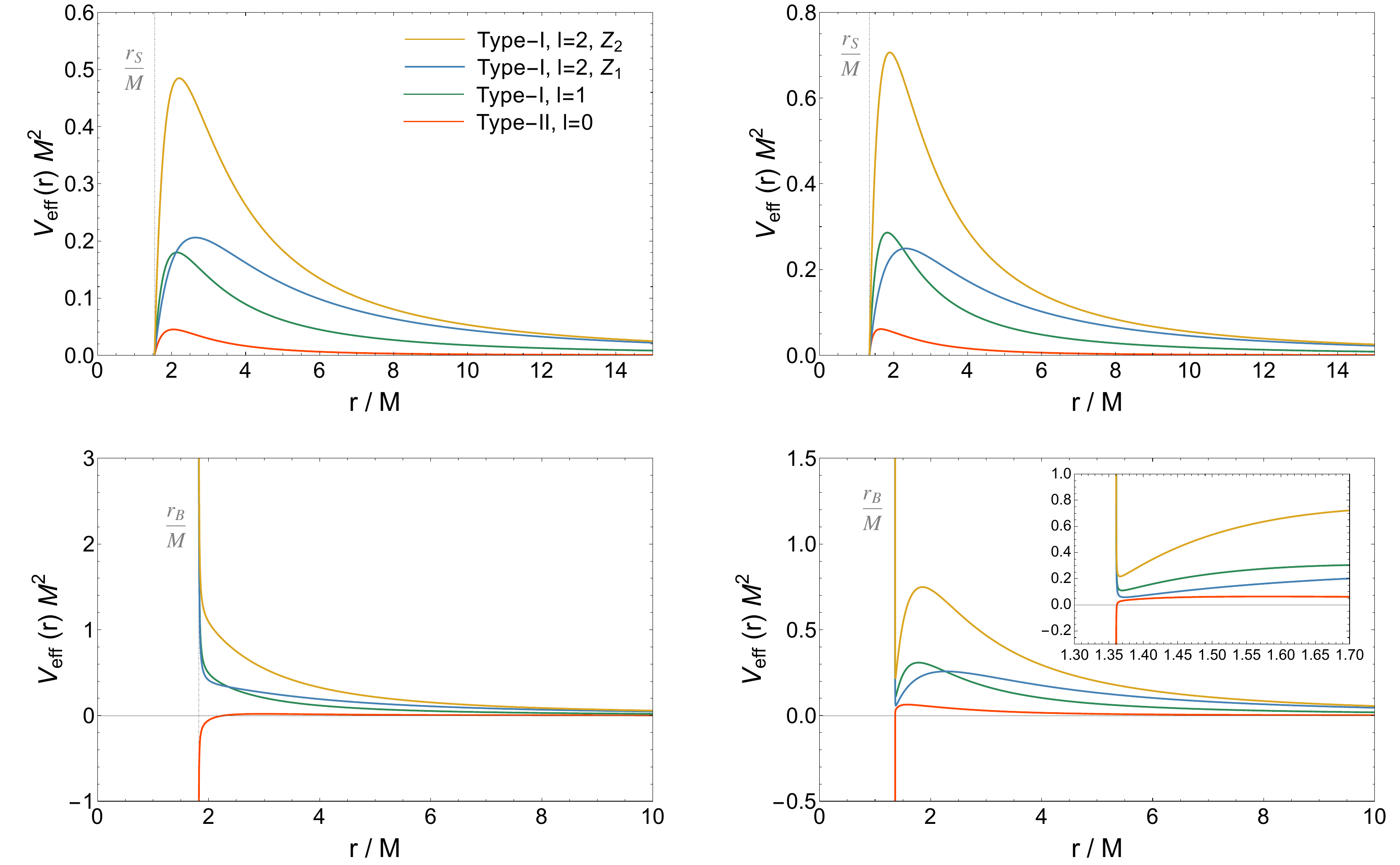}
  \caption{Effective potentials $V_\text{eff}$ (see Table~\ref{tab:potentials}) for perturbations of a magnetized BH: top left $e Q_m / M = 1.032$ (equivalently $r_B/r_S = 0.6$), top right $e Q_m / M = 1.149$ ($r_B/r_S = 0.97$); and a TS: bottom left $e Q_m / M = 1.220$ ($r_B/r_S = 1.67$), bottom right $e Q_m / M = 1.160$ ($r_B/r_S = 1.03$).} \label{fig:potentials}
\end{figure*}

\subsubsection{Comparison of effective potentials}
As shown above, the equations for the radial ($l=0$) perturbations, and for all ($l\geq1$) Type-I perturbations can be written in canonical form 
\begin{align}\label{eq:canonical}
    \frac{d^2 \Psi}{d\rho^2}+(\omega^2-V_{\rm eff})\Psi=0\,,
\end{align} 
in terms of some suitable master variable $\Psi$, generalized tortoise coordinate $\rho$, and an effective potential $V_{\rm eff}$.
For the reader's convenience, we summarize the effective potentials in Table~\ref{tab:potentials}.
\begin{table*}[]
    \centering
    \begin{tabular}{c|l}
    \hline
       \hline
       $l=0$  & $V_{\rm eff}^{l=0} = \frac{(r-r_S)\left(32r^3 -24r^2r_B - 39r_B^2r_S+rr_B(36r_S-5r_B)\right)}{16 r^5 (r-r_B)}$ \\
       $l=1$, Type-I  & $V_{\rm eff}^{l=1} = \frac{(r-r_S)\left(189r_B^4 r_S + 128 r^4 (r_B +2 r_S) + 64r^2r_B^2(6r_B+19r_S)-16 r^3r_B(27r_B+52r_S)-9rr_B^3(9r_B+92r_S)\right)}{16 r^5 (4r-3r_B)^2 (r-r_B)}$ \\
       $l\geq2$, Type-I  & $V_{\rm eff}^{(1,2)} = \frac{(r-r_S)\left(16 r^3 \Lambda - r^2 (8r_B + 24 r_S + 16 \Lambda r_B) +r(11 r_B^2 + 60 r_B r_S) - 39 r_B^2 r_S \mp 8 r (r-r_B)\sqrt{(2 r_B - 3 r_S)^2+ 12r_B r_S \Lambda}\right)}{16 r^5 (r-r_B)} $\\
       \hline
       \hline
    \end{tabular}
    \caption{Effective potentials for various types of perturbations of magnetized BHs and TSs that admit decoupled equations in Schr\"{o}dinger-like form. Note that for $l\geq2$, Type-I perturbations can be decoupled only using the explicit form of the background functions $f_S$ and $f_B$, while the other potentials can be written for a generic spherically symmetric background metric.
    }
    \label{tab:potentials}
\end{table*}

These potentials are shown in Fig.~\ref{fig:potentials} for some representative values of the parameters. We consider two magnetized BH solutions (with different values of the charge) and two TSs, of the first and second kind, respectively.

While the effective potentials in the BH case have the standard shape ~--namely they vanish at the boundaries and display a single maximum which roughly corresponds to the unstable photon sphere~-- those for perturbations of a TS have a richer structure.
First of all, it is easy to show that the effective potentials diverge at $r=r_B$.
In particular, the radial-perturbation potential diverges to \emph{negative} values. Despite this fact, we have not found any unstable mode or signature of linear instability for $r_B<2 r_S$, as later discussed.
Furthermore, the shape of the effective potentials for $l>0$ Type-I perturbations depends strongly on the background solution: only compact TSs have an unstable photon sphere at some $r>r_B$, so that they display a local maximum, a cavity, and finally a positively diverging potential at $r=r_B$. This shape of the potential naturally supports long-lived modes~\cite{Cardoso:2014sna,Heidmann:2023ojf,Bianchi:2023sfs}, as we will explicitly show below.

\subsubsection{Boundary conditions}
\label{sec:BC}

The above system of second-order differential equations is solved --~both in the frequency and in the time domain~-- imposing suitable BCs. At infinity we require radiative purely outgoing waves in all cases. If the background solution is a BH, we impose radiative purely ingoing BCs at the horizon, $r=r_S$. If the background is a TS, we impose regularity of the perturbation at the boundary $r=r_B$. Schematically, for a generic array of perturbations $\Psi$ in the frequency domain, we impose the series expansion
\begin{equation}
    \Psi = (r-r_B)^\lambda \sum_{i=0}^\infty c_i (r-r_B)^i\,, \label{eq:Frobenius}
\end{equation}
and obtain the two independent solutions by solving the indicial equation for $\lambda$. In all cases under consideration, only one of the two solutions is regular at $r=r_B$. Strictly speaking, regularity is not required in the 4D compactification as long as the 5D uplift is regular. However, regularity of the perturbations in 4D ensures also regularity in 5D, despite the fact that the 4D background is singular at $r=r_B$.
Note that this argument strictly applies only to perturbations with no excitation in the fifth dimension, which are the focus of this work. We leave the investigation of the appropriate boundary conditions for more general $y$-dependent perturbations to follow-up work in preparation~\cite{companion}.\footnote{Preliminary work that approaches the study of linear response of TSs from a full 5D perspective allows us to verify that the 5D equations indeed correctly reduce to our Type-I and radial Type-II equations in the limit of vanishing $y$-momentum. Thus, one can easily check that the regularity boundary conditions required for 5D perturbations with trivial $y$-momentum coincide with those we employ for our 4D perturbations. This is also consistent with the independent analysis of Ref.~\cite{Bena:2024hoh}, performed in 5D.}.
The coefficients $c_i$ with $i>0$ can all be written in terms of $c_0$ by solving the field equations order by order near the boundary. 
We typically solve the field equations near both boundaries perturbatively to high order, to improve numerical accuracy.

In the time domain, we perform some field/coordinate redefinition to impose the same BC. As an illustrative example, let us consider the case of Type-I dipolar perturbations.
The single equation reads
\begin{align}
&
\E''(r) 
+ \frac{1}{(r-r_B)}\left(\frac{r(2r_B+r_S)-3r_Br_S}{r(r-r_S)}\right) \E'(r) \nonumber\\
&+ \frac{1}{(r-r_B)}\left( \frac{\omega^2r^4 - 2 r^2 - 2 r (r_B-r_S) + 2r_Br_S}{r(r-r_S)^2}\right)\E(r)  = 0\,.
\end{align}
Using a series expansion as in Eq.~\eqref{eq:Frobenius}, the indicial equation is $\lambda(1+\lambda)=0$, so the two linearly independent solutions are
\begin{align}
    \E_1(r)  & =  \sum_{i=0}^\infty a_i (r-r_B)^i
    \\
    \E_2(r)  & =  (r-r_B)^{-1}\sum_{i=0}^\infty b_i (r-r_B)^i + \alpha \log(r-r_B) \E_1(r)\,.
\end{align}
The second solution is divergent at $r=r_B$. A general solution would be a linear combination of the two above which, after reabsorbing some coefficients, reads $ \E(x) =  (a_0 + a_1 x + ...) + \frac{b_0}{x} + \alpha \log(x) (c_0 + c_1 x + ...)$, with $x=r-r_B$. This suggests that the regularity condition $x\partial_x\E|_{x=0}=0$ implies correctly $b_0=0=\alpha c_0$.

For the Type-II radial equation, one obtains two identical roots of the indicial equation, $\lambda^2=0$. The most general solution in the asymptotic limit $r\rightarrow r_B$ can be written as
\begin{align}
    \varphi(r)  & = \sum_{i=0}^\infty a_i (r-r_B)^i + \log(r-r_B) \sum_{i=0}^\infty b_i (r-r_B)^i
\end{align}
As explained before, we can impose  $x\partial_x\varphi|_{x=0}=0$ to ensure regularity of the solution.

A similar procedure applies to any kind of perturbations, including those in coupled systems.

\section{Spectroscopy of magnetized BHs and TSs} \label{sec:results}
This section presents our numerical results for the linear perturbations of magnetized BHs and TSs. Besides the technicalities related to the coupling between scalar, EM, and gravitational perturbations of different parities, the spectrum of magnetized BHs is standard and qualitatively similar to that of a Reissner-Nordstr\"{o}m BH.
The spectrum of TSs is instead richer and strongly depends on the parameter space of the background solution. TSs of the first kind do not have a stable photon ring and their QNMs are similar to the gravitational $w$-modes of compact stars in general relativity~\cite{Kokkotas:1999bd} and hence qualitatively similar to BHs.
TSs of the second kind have a pair of stable and unstable photon spheres, which can support long-lived QNMs, as expected for ultracompact objects~\cite{Cardoso:2019rvt}.
Their response in the time domain is initially very similar to that of a BH with comparable charge-to-mass ratio, while their late time response is governed by echoes~\cite{Cardoso:2016oxy,Cardoso:2016rao,Cardoso:2017cqb}, analogously to what observed for test scalar perturbations of microstate geometries~\cite{Ikeda:2021uvc}.

\subsection{Numerical methods}
We have computed the QNMs of magnetized BHs and TSs both in the frequency domain, solving an eigenvalue problem, and in the time domain, solving a $1+1$ evolution problem and then extracting the QNMs from the inverse Fourier transform of the signal. As discussed below, the two complementary methods show excellent agreement. In the next two subsections we shall discuss some details of the numerical implementation.

\subsubsection{Frequency-domain computations}

The frequency-domain analysis is performed by computing the QNMs of both magnetized BHs and TSs employing a direct integration shooting method \cite{Pani:2013pma,Rosa:2011my,Ferrari:2007rc,Pani:2012bp}. At the boundary $r = r_B$ we expand the perturbation as in Eq.~\eqref{eq:Frobenius} imposing regularity (the same holds for magnetized BHs in the vicinity of the horizon at $r = r_S$, imposing purely ingoing waves). Solving the equations near the boundaries as a Frobenius series one can obtain the Frobenius index $\lambda$ and the coefficients $c_i$ with $i > 0$ in terms of $c_0$. We can set $c_0=1$ without loss of generality using the fact that the perturbation equations are linear.
For coupled systems of equations the initial conditions are generically $N$-dimensional and one can choose an orthogonal basis $c_0=(1,0,0,..,0)$, $c_0=(0,1,0,...,0)$, ..., $c_0=(0,0,0,...,1)$ of $N$ dimensional unit vectors as explained in~\cite{Pani:2013pma}.
We can then use the BC at the inner boundary to integrate the radial equations up to arbitrarily large distance. Asymptotically at infinity we expect the general solution to be a linear combination of an ingoing and an outgoing wave
\begin{align}
    \Psi \sim B(\omega)\, e^{- i \omega r}\, r^{\Tilde{\lambda}} +  C(\omega)\, e^{i \omega r}\, r^{-\Tilde{\lambda}} \hspace{0.3cm} r \rightarrow +\infty \,,
\end{align}
where $B(\omega)$ and $C(\omega)$ are complex coefficients and the exponent $\Tilde{\lambda}$ can be derived by solving the equations order by order. In order to find the discrete spectrum of complex frequencies of the QNMs we impose the asymptotic BC of purely outgoing waves, namely we require $B (\omega) = 0$. (For an $N$-dimensional problem, this condition generalizes to the vanishing of a determinant obtained from the $N$-dimensional basis.)
Since $B$ is a complex function of $\omega$, imposing $B (\omega) = 0$ is achieved through a shooting procedure in the complex $\omega$ plane. All the computations discussed above are performed using \textsc{Mathematica} with large numerical precision and high-order series expansions at the boundary.

\subsubsection{Time-domain computations}

In parallel with the frequency-domain approach, we solve numerically the time evolution equations of linear perturbations and extract the QNMs using spectral analysis techniques. Specifically, by inverse-Fourier transforming the canonical equation~\eqref{eq:canonical}, we consider

\begin{align}\label{eq:canonicaltime}
   \left[\frac{d^2 }{d\rho^2}-\frac{d^2 }{dt^2}-V_{\rm eff}\right]\Psi(t,\rho)=0\,.
\end{align} 

To conduct the time evolution we resort to a custom PDE solver written in \Julia~\cite{bezanson2017julia}, based on algorithms collected in the \DiffEq ~suite~\cite{rackauckas2017differentialequations}, which is part of the \SCIML ~library of Open Source Software for Scientific Machine Learning. The numerical method we adopt is based on the method of lines: we approximate the spatial derivatives with a standard fourth-order finite difference stencil and employ a fourth-order Runge-Kutta algorithm for the time stepping. The boundary treatment consists in fourth-order finite difference approximations of physical BCs. For magnetized BHs, these consist in ingoing/outgoing radiative conditions, respectively at the inner/outer boundary. Instead, as discussed in the previous section~\ref{sec:BC}, TSs require the ingoing BC to be replaced by regularity conditions at the surface $r=r_B$, corresponding to $(r-r_B) \partial_r X \rightarrow 0$, where $X$ is to be replaced with the appropriate linear perturbation.

As a relevant technical detail, we report that we solved the time evolution equations expressed in the standard radial tortoise coordinate, $dr_* = f_S^{-1} dr$, instead of the coordinate radius $r$ or the generalized tortoise coordinate $\rho$ defined in Eq.~\eqref{eq:gentortoise}. Compared to the latter, the choice of $r_*$ allows us to avoid dealing with terms $\sim f_B^{-1}$ in the effective potentials, which would require regularization in a neighborhood of $r=r_B$ (equivalently, $\rho=0$) to avoid spurious numerical instabilities. At the same time, in the case of TSs perturbations, a grid in $r_*$ allows having a resolution that increases like $\sim (r-r_S)^{-1}$ near the surface of the star. Empirically, this turns out to be crucial to properly resolve the cavity effects in TSs of the second kind. 

In addition, we find that we need long simulations (i.e., up to $t/M \sim {\cal O}( 10^3\textup{--}10^4)$) to be able to extract the long-lived modes with damping timescale $ \sim {\cal O}(10^5)\textup{--} {\cal O}(10^{10})$ with sufficient accuracy. This was made possible by adopting an ad-hoc coordinate stretching, that allowed us to push the outer boundary sufficiently far away from the effective potential as to not contaminate the simulation with spurious boundary effects. 

To estimate the QNM frequencies in our time evolution simulations, we produce a time series by extracting the amplitude of each linear perturbation field at an arbitrary fixed point, typically corresponding to $r_{\rm ex}=20 M$. We then process this signal by windowing and applying a Fast Fourier Transform to determine the corresponding spectrum of the perturbations. Then, we move to fitting each peak in two steps: first, we employ a nonlinear fit using a Lorentzian function to have an estimate of the real part of the frequency, $\omega_R$. Unfortunately, this step does not yield an equally accurate estimate of the imaginary part of the QNM frequency. To obviate this inconvenience, we filter the original time series in such a way as to suppress all modes with frequencies that do not match the estimated $\omega_R$. Then, we apply a nonlinear fit with a damped sinusoid template to the filtered signal. This provides us with a refined estimate of $\omega_I$. An example of the power spectrum will be discussed in Sec.~\ref{sec:time} below. For further details on this spectral analysis technique, we refer the interested reader to previous applications to similar problems~(e.g., ~\cite{Dolan:2012yt}).

Our implementation has been validated by comparing results against tabulated Schwarzschild QNMs~\cite{Berti:2009kk} and cross-checking between the frequency-domain and time-domain frameworks. In addition, we discuss numerical convergence in Appendix~\ref{app:convergence}.

\begin{figure*}[th]
  \centering
  \includegraphics[width=1\textwidth]{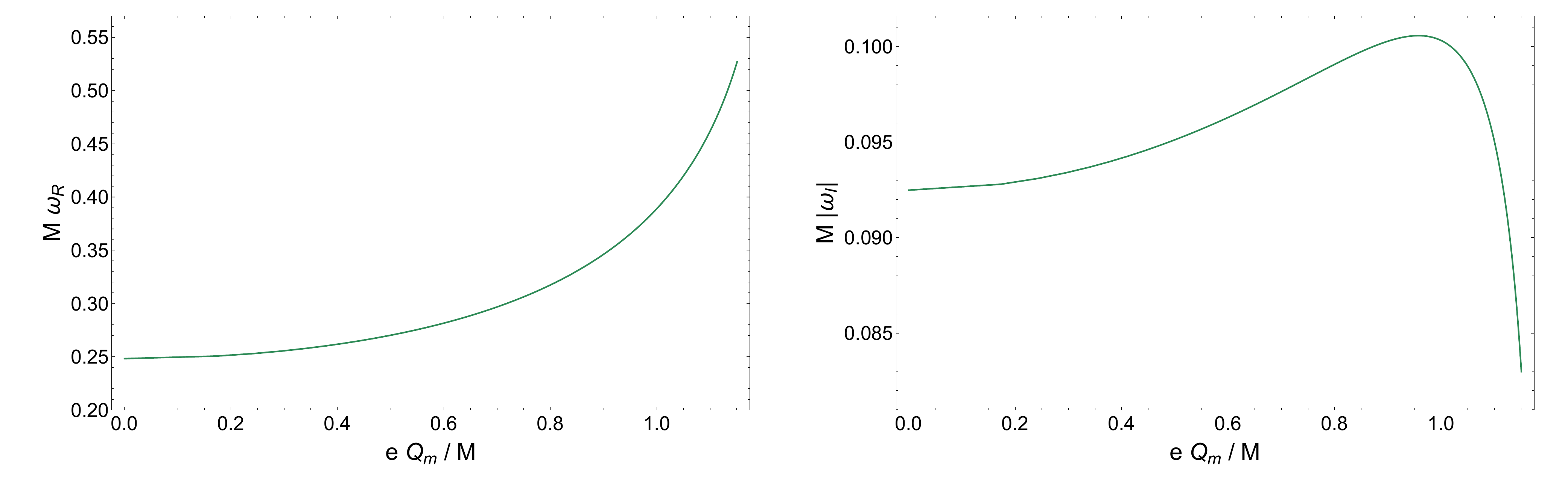}\\
  \includegraphics[width=1\textwidth]{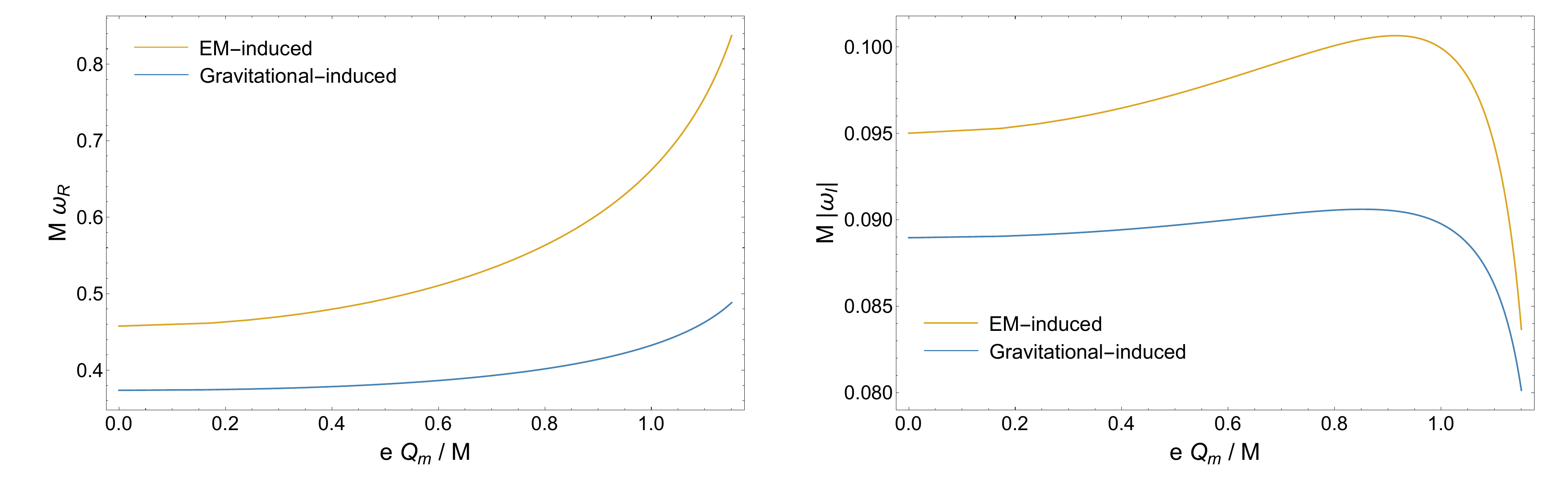}\\
  \includegraphics[width=1\textwidth]{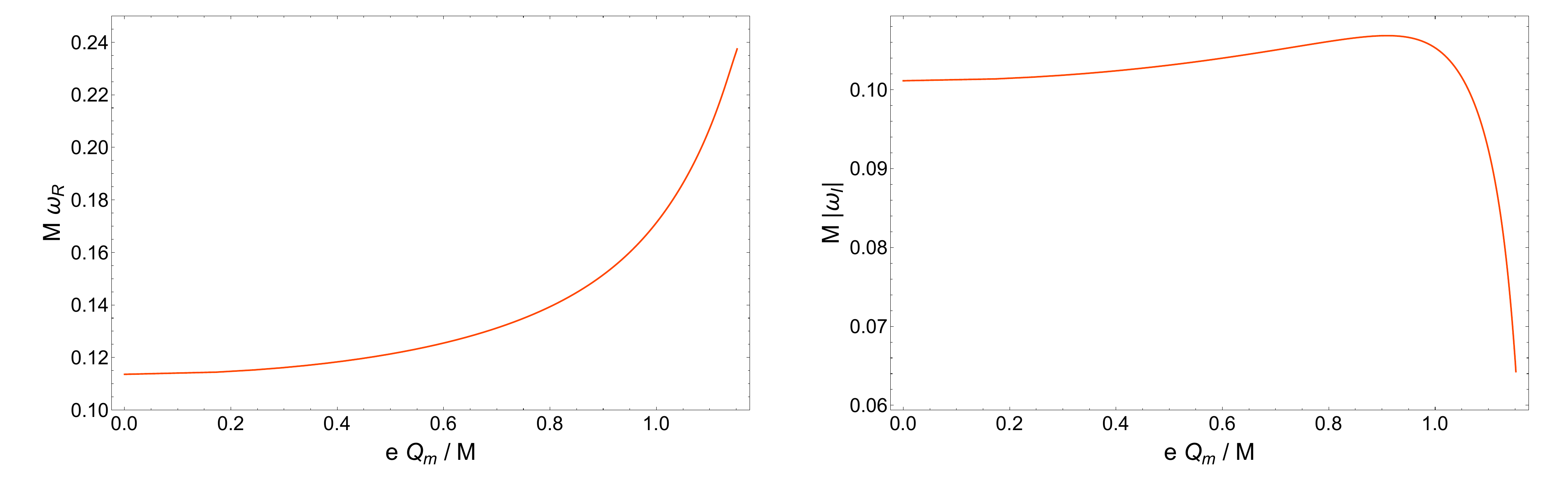}
  \caption{Fundamental QNMs of a magnetized BH as a function of the magnetic charge. The left (right) panels show the real (imaginary) part of the mode.
  Top panels: even EM perturbations (Type-I, $l=1$).
  Middle panels: even EM and odd metric perturbations (Type-I, $l=2$).
  Bottom panels: scalar perturbations (Type-II, $l=0$).
  }\label{fig:BHtypeIandl0}
\end{figure*}

\begin{figure*}[th]
  \centering
  \includegraphics[width=1\textwidth]{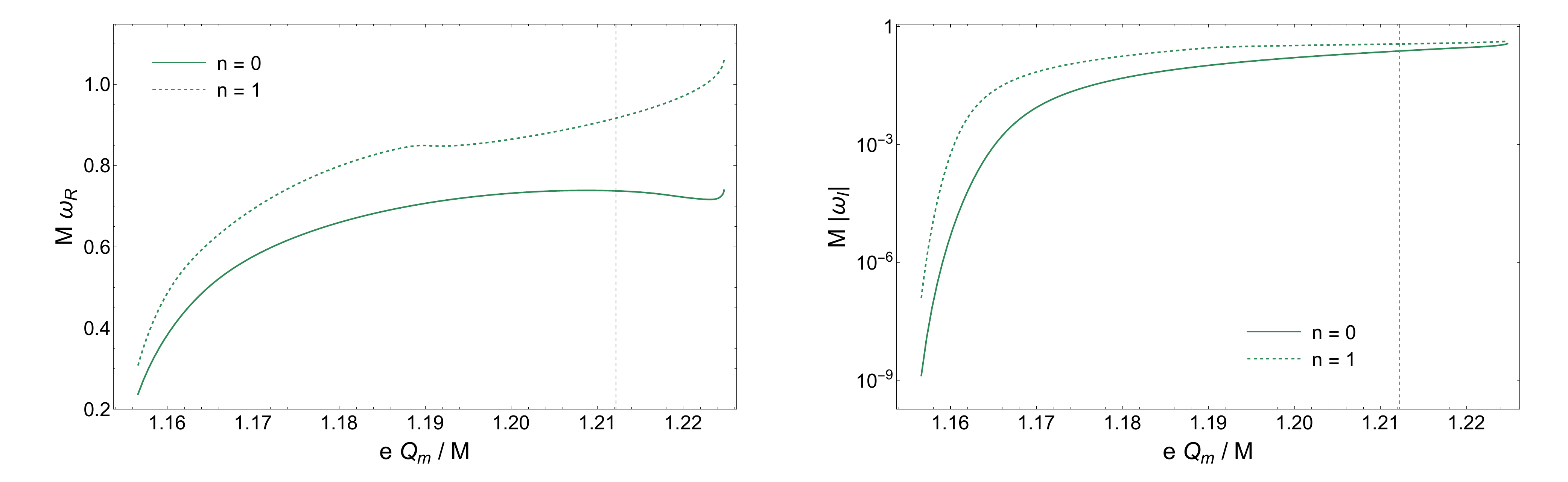}\\
    \includegraphics[width=1\textwidth]{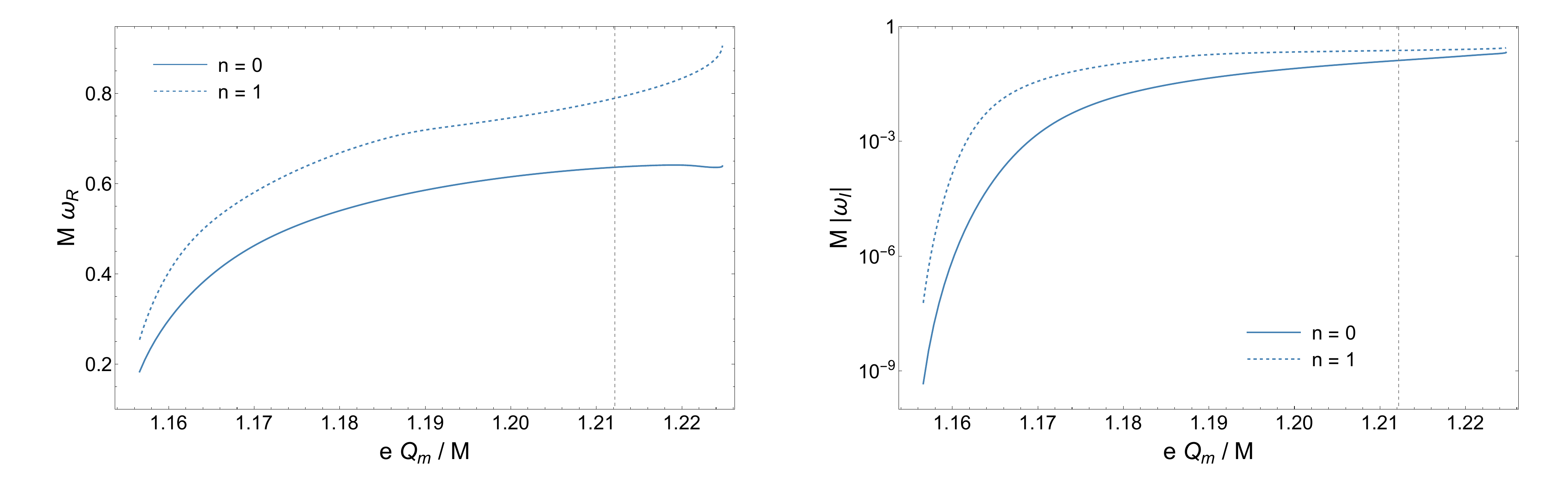}\\
  \includegraphics[width=1\textwidth]{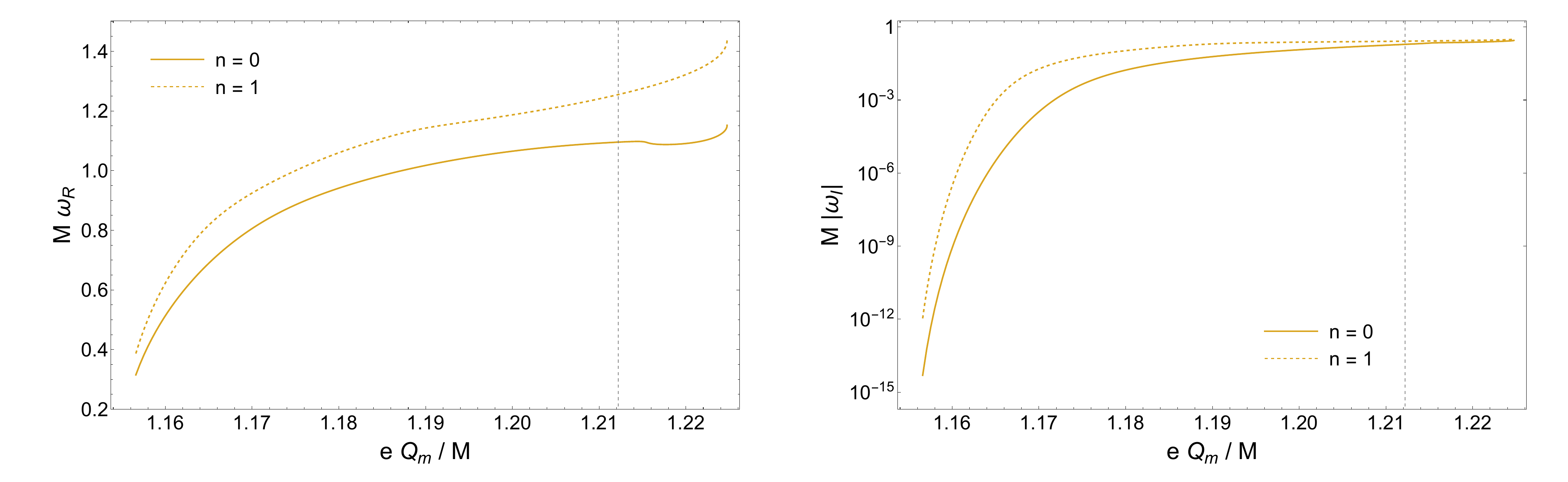}\\
  \includegraphics[width=1\textwidth]{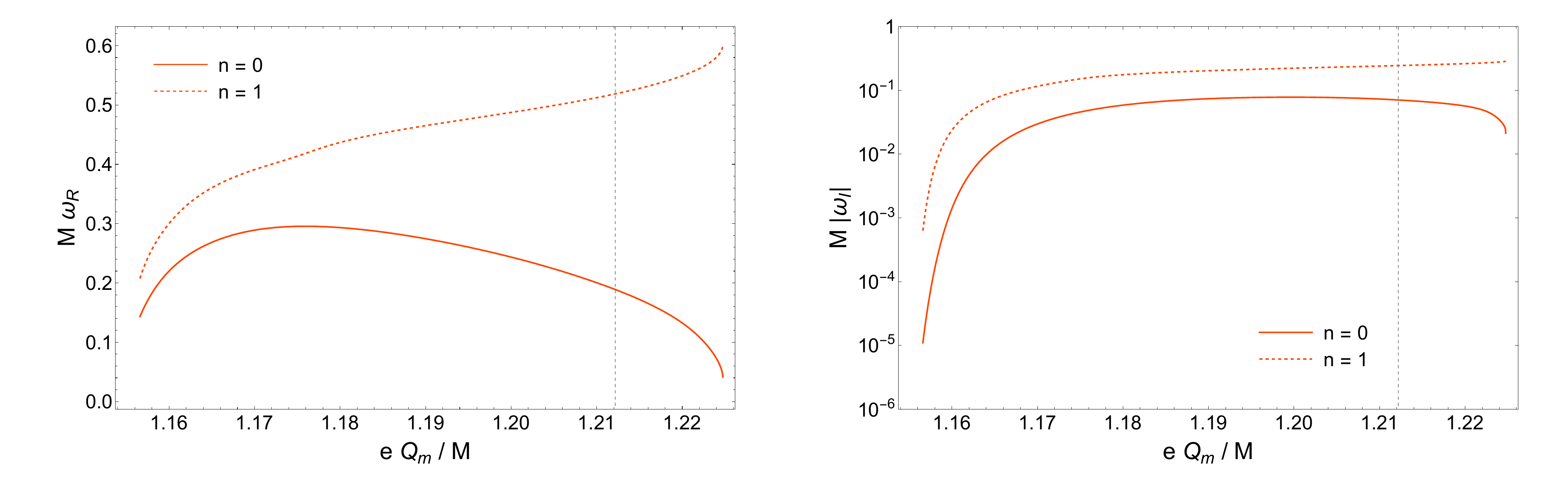}
  \caption{Same as Fig.~\ref{fig:BHtypeIandl0} but for TSs. Note that in this case the modes smoothly interpolate between being long-lived for a TS of the second kind to BH-like modes for a TS of the first kind (the vertical dashed line denotes the transition between the two regimes). First panels: even EM perturbations (Type I, $l=1$). Second and third panels: odd metric and even EM perturbations (Type I, $l=2$) respectively. Fourth panels scalar perturbations (Type II, $l=0$).
}\label{fig:TStypeIandl0}
\end{figure*}

\begin{table*}[h!]
    \centering
    \begin{tabular}{|c|c|c|c|c|}
     \hline
  \multicolumn{2}{|c}{} \vline  & {\bf Magnetized BH} & {\bf TS, second kind} & {\bf TS, first kind} \\
    \hline
  \multirow{2}{*}{$n=0$}  & f-domain  & $ 0.530286 - \I 8.207 \times 10^{-2}$  &  $0.237163 - \I 1.326 \times 10^{-9}$ &  $0.752068 - \I 0.248$ \\
   \cline{2-5}
    & t-domain  & $ 0.530486 - \I 8.140 \times 10^{-2}$ & $0.237247 - \I 1.412 \times 10^{-9}$ &  $0.759523 - \I 0.238$ \\
   \hline
  \multirow{2}{*}{$n=1$}  & f-domain  & - & $0.30767\textcolor{white}{0} - \I 1.24\textcolor{white}{0} \times 10^{-7}$ &  - \\ 
    \cline{2-5}
    & t-domain  & - & $ 0.307681 - \I 1.240 \times 10^{-7}$ & - \\
    \hline
  \multirow{2}{*}{$n=2$}  & f-domain  & - & $0.376555 - \I 4.521 \times 10^{-6}$ & - \\
   \cline{2-5}
    & t-domain & -  & $0.376595 - \I 4.532 \times 10^{-6}$ & - \\
    \hline
  \multirow{2}{*}{$n=3$}  & f-domain  & - &  $ 0.443242 - \I 9.362 \times 10^{-5} $ & - \\
    \cline{2-5}
    & t-domain &  - & $0.44339\textcolor{white}{0} - \I 9.34\textcolor{white}{0} \times 10^{-5}$ & - \\
    \hline
  \multirow{2}{*}{$n=4$}  & f-domain  & - & $0.506573 - \I 1.141 \times 10^{-3}$ & - \\
      \cline{2-5}
      & t-domain &   - &$0.50653\textcolor{white}{0} - \I 1.41\textcolor{white}{0}  \times 10^{-3}$ & -\\
      \hline
  \multirow{2}{*}{$n=5$}  & f-domain  & - &  $0.566991 - \I 6.682 \times 10^{-3}$ & - \\
      \cline{2-5}
      & t-domain & - &  $0.56660\textcolor{white}{0} - \I 6.950  \times 10^{-3}$ & -\\
      \hline
        \multirow{2}{*}{$n=6$}  & f-domain  & - & $0.630407 - \I 1.955 \times 10^{-2}$ & -\\
      \cline{2-5}
      & t-domain & - & $0.6269\textcolor{white}{00} - \I 2.28\textcolor{white}{0} \times 10^{-2}$ & - \\
      \hline
    \end{tabular}
    \caption{Type-I QNMs, $l=1$: magnetized BH with $e Q_m/M=1.153$ (equivalently, $r_B/r_S=0.99$), second kind TS with $e Q_m/M=1.157$ ($r_B/r_S=1.01$), and first kind TS with $e Q_m/M=1.220$  ($r_B/r_S=1.67$). For the magnetized BH and first-kind TS we computed only the fundamental mode ($n=0$), while for the second-kind TS we computed also the first overtones ($n=1,2,3,...$). In all cases we compare the QNMs computed in the frequency domain with those extracted from the Fourier transform of the time domain signal. 
    Here and in subsequent tables, the QNMs are normalized by the mass, i.e. we show the complex quantity $M\omega$, and - means that the mode has not been computed.
    }
\label{tab:TypeI_l1_QNMs}
\end{table*}

\subsection{Type-I QNMs of magnetized BHs and TSs}
We start presenting the Type-I perturbations, which are less involved than the Type-II case. Indeed, the Type-I sector couples odd-parity gravito-scalar perturbations with even-parity EM perturbations. Since scalar perturbations have even parity and odd-parity gravitational perturbations are easier than their even-parity counterpart, in the Type-I sector we have fewer degrees of freedom and no $l=0$ modes.

We present the QNMs in the form $\omega =\omega_R +\I \omega_I $, typically normalizing them by the mass, i.e. $\omega M$. Given our conventions, $\omega_I<0$ (resp., $\omega_I>0$) corresponds to a stable (resp., unstable) mode.

\subsubsection{Dipolar perturbations}
Dipolar ($l=1$) Type-I perturbations are described by a single master variable governed by Eq.~\eqref{eq:E_B&Hcoord_l1}.

In Table~\ref{tab:TypeI_l1_QNMs} we show the QNMs for some representative examples of magnetized BHs and TSs. In particular we consider a nearly-extremal magnetized BH with $e Q_m/M\approx 1.153$, a TS of the first kind with $e Q_m/M\approx 1.220$, and a TS of the second kind with $e Q_m/M\approx 1.157$, so with a charge-to-mass ratio very similar to that of the BH.
We show results obtained both in the frequency and in the time domain using the methods previously discussed. As evident from this and similar tables presented below, the agreement of the two methods is very good, even for higher-order overtones. QNMs with smaller quality factor, $\omega_R/|\omega_I|$, are less accurate, because in this case the direct integration method is less efficient and the accuracy of the power spectrum extracted from the time-domain signal is limited by the short duration of the mode.

For BHs and first-kind TSs we show only the fundamental mode, which already has a relatively short damping time. For second-kind TSs, we find long-lived modes (with $|\omega_I|\ll\omega_R$), as expected. In this case we computed several overtones up to a point in which their imaginary part is comparable to that of an ordinary BH QNM.

The fundamental QNM of magnetized BHs in this sector is shown in the top panels of Fig.~\ref{fig:BHtypeIandl0} as a function of $Q_m/M$ up to the nearly-extremal case.
The behavior of this mode with the BH charge is qualitatively similar to that  of a Reissner-Nordstr\"{o}m BH (see, e.g.,~\cite{Berti:2009kk}).
More interestingly, the top panels of Fig.~\ref{fig:TStypeIandl0} track the behavior of some Type-I dipolar QNMs of the TS as a function of $Q_m/M$, in particular across the smooth transition between first- and second-kind solutions.
As expected, we see that  first-kind solutions have BH-like modes, with imaginary part only slightly smaller than the real one, which are akin to the so-called $w$-modes of a neutron star~\cite{Kokkotas:1999bd}. However, as the charge-to-mass ratio decreases, the solution develops a stable photon sphere and the mode becomes long-lived, as expected for a ultracompact object. We track the fundamental mode ($n=0$) and the first overtone ($n=1$). Interestingly, the damping time of the latter is longer than that of the former for any $Q_m$ (i.e., the curves on the right panel do not cross each other), so the fundamental mode does not change during the tracking at different $Q_m$.

\begin{figure*}[th]
  \centering
  \includegraphics[width=1\textwidth]{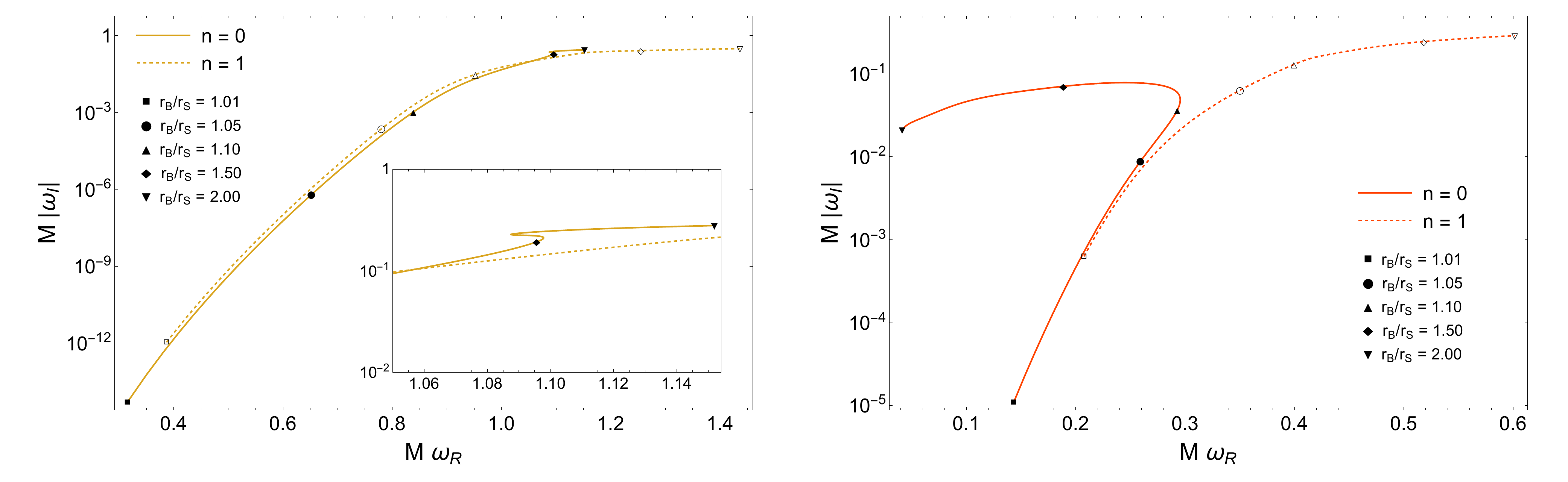}
  \caption{Fundamental QNMs for TSs of even EM perturbations (Type I, $l=1$, left panel) and scalar perturbations (Type II, $l = 0$, right panel).
}\label{fig:TS_ImvsReal}
\end{figure*}

For completeness, in Fig.~\ref{fig:TS_ImvsReal} we track the modes for different values of $Q_m/M$ in the complex $(\omega_R,\omega_I)$ plane, where the transition between first- and second-kind TSs is more evident.

\subsubsection{$l\geq2$ perturbations}
Type-I perturbations with $l\geq2$ are described by the two decoupled master equations~\eqref{eq:typeIdecoupled} but gravitational and EM perturbations are anyway mixed. In the decoupling limit ($r_B\to0$), the master variables $Z_1$ and $Z_2$ are associated to gravitational and EM perturbations of a Schwarzschild BH, respectively. As such, we will refer to modes coming from the first and second equation in~\eqref{eq:typeIdecoupled} as \emph{gravitational-induced} and \emph{EM-induced}, respectively, even for generic values of $r_B/r_S$ where an actual decoupling of the original perturbations is not possible.

Examples of the QNMs in this sector for BHs and TSs are shown in Table~\ref{tab:TypeI_l2_QNMs} and Table~\ref{tab:TypeI_l2_QNMsEM} for the gravitational-induced and EM-induced modes, respectively.
Some modes of both families are tracked as function of $Q_m/M$ in the middle panels of Fig.~\ref{fig:BHtypeIandl0} and Fig.~\ref{fig:TStypeIandl0} for BHs and TSs, respectively.
Beside this doubling of modes, we observe the same qualitative behavior as previously discussed for $l=1$ Type-I modes.

\begin{table*}[h!]
    \centering
    \begin{tabular}{|c|c|c|c|c|}
    \hline
    \multicolumn{2}{|c}{} \vline  & {\bf Magnetized BH} & {\bf TS, second kind} & {\bf TS, first kind} \\
    \hline
    \multirow{2}{*}{$n=0$}  & f-domain  & $0.489568 - \I 7.972 \times 10^{-2}$ &  $0.183217 - 
    \I 4.674 \times 10^{-10}$ &  $ 0.644348 - \I 0.1551 $ \\
    \cline{2-5}
    & t-domain  & $0.489600 - \I 7.978 \times 10^{-2}$ & $0.183219 - \I 3.349 \times 10^{-10} $ &  $0.643938 - \I 0.1665 $ \\
    \hline
    \multirow{2}{*}{$n=1$}  & f-domain  & -  &  $ 0.254071 - \I 6.001 \times 10^{-8}$ &  - \\
    \cline{2-5}
    & t-domain  & - & $0.254084 - \I 6.008 \times 10^{-8}$ &  - \\
    \hline
    \multirow{2}{*}{$n=2$}  & f-domain  & -  &  $0.323219 - \I 2.615 \times 10^{-6}$ &  - \\
    \cline{2-5}
    & t-domain  & - & $0.323263 - \I 2.622 \times 10^{-6}$ &  - \\
    \hline
    \multirow{2}{*}{$n=3$}  & f-domain  & -  &  $0.390169 - \I 6.116 \times 10^{-5}$ &  - \\
    \cline{2-5}
    & t-domain  & - & $0.390256 - \I 6.142 \times 10^{-5}$ &  - \\
    \hline
    \multirow{2}{*}{$n=4$}  & f-domain  & -  &  $0.453786  - \I 8.348 \times 10^{-4}$ &  - \\
    \cline{2-5}
    & t-domain  & - & $0.453832 - \I 8.340 \times 10^{-4}$ &  - \\
    \hline
    \multirow{2}{*}{$n=5$}  & f-domain  & -  & $0.513765  - \I 5.463 \times 10^{-3}$ &  - \\
    \cline{2-5}
    & t-domain  & - & $0.513375 - \I 2.754 \times 10^{-3}$ &  - \\
    \hline
    \multirow{2}{*}{$n=6$}  & f-domain  & -  & $0.574947 - \I 1.658 \times 10^{-2}$ &  - \\
    \cline{2-5}
    & t-domain  & - & $0.572869  - \I 1.140 \times 10^{-2}$ &  - \\
    \hline
    \end{tabular}
    \caption{Same as Table~\ref{tab:TypeI_l1_QNMs} but for $l=2$, gravitational-induced Type-I perturbations.    
    }
    \label{tab:TypeI_l2_QNMs}
\end{table*}

\begin{table*}[h!]
    \centering
    \begin{tabular}{|c|c|c|c|c|}
    \hline
    \multicolumn{2}{|c}{} \vline  & {\bf Magnetized BH} & {\bf TS, second kind} & {\bf TS, first kind} \\
    \hline
    \multirow{2}{*}{$n=0$}  & f-domain  & $ 0.841470 - \I 8.294 \times 10^{-2}$ &  $0.315245 - \I 4.949 \times 10^{-15}$ &  $1.09087 - \I 0.1836$ \\
    \cline{2-5}
    & t-domain  & $0.841302 - \I 8.296 \times 10^{-2}$ & $0.315258 - \I  (*) \textcolor{white}{00\times 10^{-00}}$ &  $1.09109 -\I 0.2110$ \\
    \hline
    \multirow{2}{*}{$n=1$}  & f-domain  & - &  $0.386772 - \I 1.074 \times 10^{-12}$ &  - \\
    \cline{2-5}
    & t-domain  & - & $0.386809 - \I (*) \textcolor{white}{00\times 10^{-00}}$ &  - \\
    \hline
    \multirow{2}{*}{$n=2$}  & f-domain  & - &  $0.457413  - \I 8.441 \times 10^{-11}$ &  - \\
    \cline{2-5}
    & t-domain  & - & $0.457631 - \I (*) \textcolor{white}{00\times 10^{-00}}$ &  - \\
    \hline
    \multirow{2}{*}{$n=3$}  & f-domain  & - &  $ 0.527134 - \I 3.691 \times 10^{-9}$ &  - \\
    \cline{2-5}
    & t-domain  & - & $0.527232 - \I 3.697 \times 10^{-9}$ &  - \\
    \hline
    \multirow{2}{*}{$n=4$}  & f-domain  &  - &  $ 0.596474 - \I 1.133 \times 10^{-7}$ &  - \\
    \cline{2-5}
    & t-domain  & - & $0.595908 - \I 1.077 \times 10^{-7} $ &  - \\
    \hline
    \multirow{2}{*}{$n=5$}  & f-domain  &  -   &  $ 0.668166 - \I 3.204 \times 10^{-6}$ &  - \\
    \cline{2-5}
    & t-domain  & - & $0.663237 - \I 2.284 \times 10^{-6}$ &  - \\
    \hline
    \end{tabular}
    \caption{Same as Table~\ref{tab:TypeI_l2_QNMs} but for EM-induced perturbations. The $(*)$ in the t-domain rows of the fundamental and first two overtones indicate that in these cases it was not possible to estimate the immaginary part of the frequency with sufficient accuracy.
    }
    \label{tab:TypeI_l2_QNMsEM}
\end{table*}

\subsection{Type-II QNMs of magnetized BHs and TSs: radial case}
Let us now turn our attention to the more involved case of Type-II perturbations, which couple even-parity gravito-scalar perturbations with odd-parity EM perturbations. Also scalar perturbations are excited in this case, starting from the monopolar ($l=0$) perturbations.

Perturbations with $l\geq2$ and with $l=1$ involve three and two propagating degrees of freedom, respectively, and the resulting system of equations does not appear to be diagonalizable. We postpone the numerical analysis of Type-II perturbations with $l\geq1$ to a companion paper~\cite{companion}. Here we focus on the case of radial perturbations.

Radial ($l=0$) perturbations only exist in the Type-II sector since in this case only the (even-parity) scalar perturbations are dynamical and described by Eq.~\eqref{eq:typeII_phic}. This sector is particularly interesting because, as shown in Fig.~\ref{fig:potentials}, the effective potential for TSs is negative and divergent as $r\to r_B$, which might signal an instability in the spectrum, i.e. QNMs with positive imaginary part.

We have searched for unstable modes and did not find any for $r_B<2r_S$, also in agreement with the time evolution presented below (see Sec.~\ref{sec:time}) which does not show any evidence for an instability.
An example of the radial QNMs of magnetized BHs and TSs is presented in Table~\ref{tab:TypeII_l0_QNMs}. The fundamental mode of magnetized BHs and the $n=0,1$ modes of TSs as a function of the charge-to-mass ratio are presented in the bottom panels of Figs.~\ref{fig:BHtypeIandl0} and~\ref{fig:TStypeIandl0}, respectively.

\begin{figure}[th]
  \centering
  \includegraphics[width=0.46\textwidth]{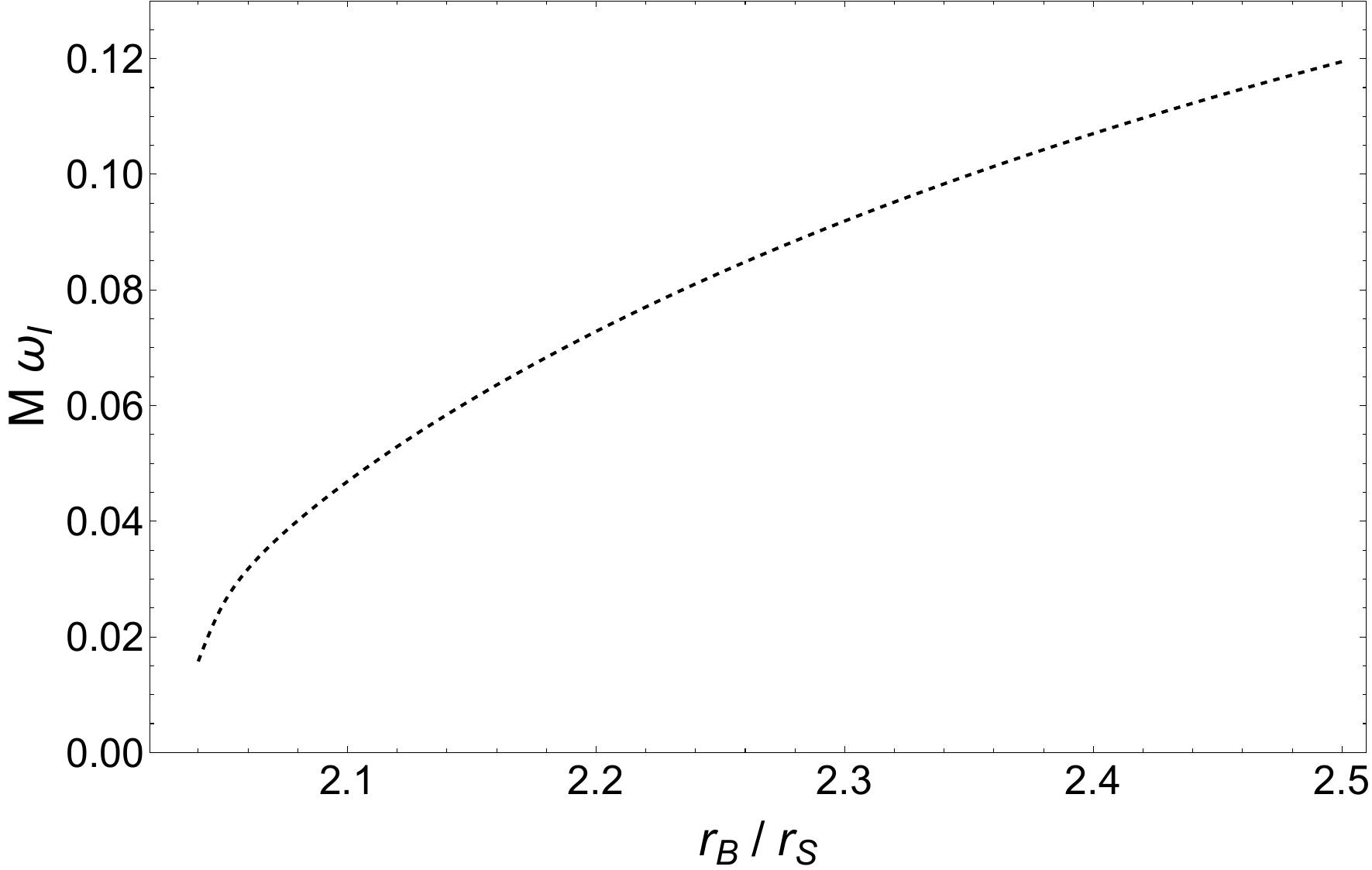}
  \caption{Purely imaginary frequency of the unstable mode of a TS as a function of $r_B/r_S$ under scalar perturbations (Type II, $l=0$). This instability exists only for $r_B>2r_S$.}\label{fig:TStypeIIandl0unstable}
\end{figure}

Finally, as previously discussed, TSs with  $r_B > 2 r_S$ are unstable under radial perturbations with purely imaginary frequency, as a consequence of the Gregory-Laflamme instability of magnetized black strings with $r_B \leq r_S/2$ and the duality $(t, y, r_S , r_B) \to (iy, it, r_B , r_S)$ that maps magnetized black strings to TSs and viceversa~\cite{Bah:2021irr}. 
In agreement with the analysis in~\cite{Bah:2021irr}, for TSs with $r_B>2r_S$ we found an unstable purely imaginary mode, i.e. $\omega=i\omega_I$ with $\omega_I>0$, for radial perturbations with zero Kaluza-Klein momentum. As shown in Fig.~\ref{fig:TStypeIIandl0unstable}, the frequency approaches zero in the $r_B = 2 r_S$ limit, thus reaching the threshold of the Gregory-Laflamme zero mode of the corresponding black string.

\begin{table*}[h!]
    \centering
    \begin{tabular}{|c|c|c|c|c|}
     \hline
  \multicolumn{2}{|c}{} \vline  & {\bf Magnetized BH} & {\bf TS, second kind} & {\bf TS, first kind} \\
    \hline
  \multirow{2}{*}{$n=0$}  & f-domain  & $ 0.237009 - \I 6.249 \times 10^{-2}$  &  $0.143401 - \I 1.113 \times 10^{-5}$ &  $0.135919 - \I 5.716 \times 10^{-2}$ \\
   \cline{2-5}
    & t-domain  & $0.237393 - \I 5.953 \times 10^{-2}$ & $0.143224 - \I 1.092 \times 10^{-5}$&  $0.133896 - \I 2.986 \times 10^{-2}$ \\
   \hline
  \multirow{2}{*}{$n=1$}  & f-domain  & - & $0.208064  - \I 6.541 \times 10^{-4}$ &  - \\ 
    \cline{2-5}
    & t-domain  & - & $0.207059 - \I 6.065 \times 10^{-4}$ & - \\
    \hline
  \multirow{2}{*}{$n=2$}  & f-domain  & - & $0.266300  - \I 7.071 \times 10^{-3}$ & - \\
   \cline{2-5}
    & t-domain & -  & $0.263869 - \I 6.273 \times 10^{-3}$ & - \\
    \hline
  \multirow{2}{*}{$n=3$}  & f-domain  & - &  $ 0.325630 - \I 2.225 \times 10^{-2} $ & - \\
    \cline{2-5}
    & t-domain &  - & $0.322313 - \I 2.612 \times 10^{-2}$ & - \\
    \hline
    \end{tabular}
    \caption{
    Same as Table~\ref{tab:TypeI_l1_QNMs} but for radial (Type-II, $l=0$) perturbations.   
    }
    \label{tab:TypeII_l0_QNMs}
\end{table*}

\subsection{Time signal and comparison between magnetized BHs and TSs} \label{sec:time}
In the previous sections we have compared the results of QNM computation in the frequency domain with those extracted from the inverse Fourier transform of the signal in the time domain, the latter being obtained by evolving a system of $1+1$ equations.
Here we present the results of the time-domain analysis.

\begin{figure}[th]
  \centering
  \includegraphics[width=0.475\textwidth]{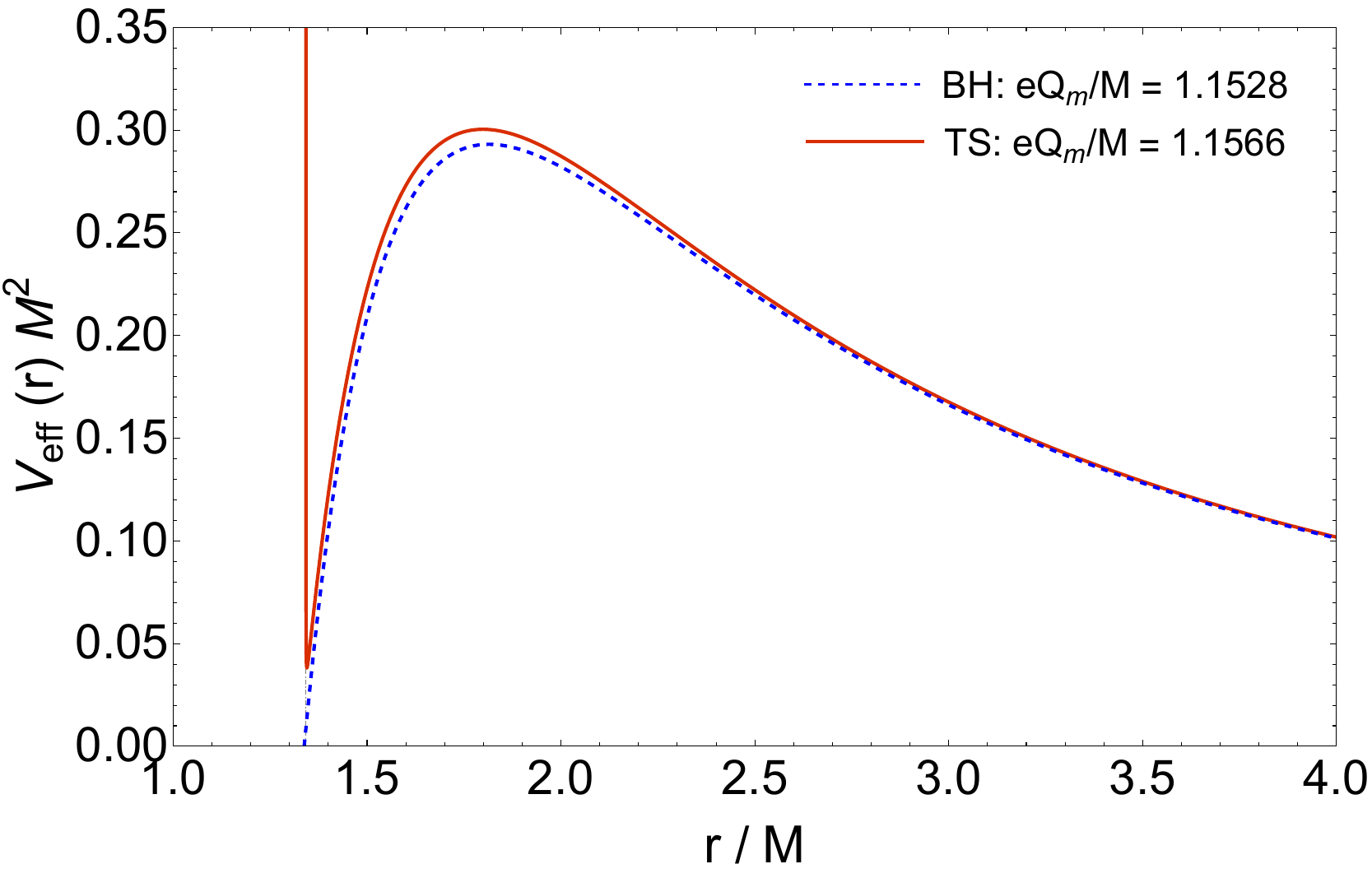}
  \caption{Comparison between the effective potentials for $l=1$ Type-I perturbations of a magnetized BH with $eQ_m/M\approx 1.1528$ and a TS with $eQ_m/M\approx 1.1566$.
 } \label{fig:potentialBHTS}
\end{figure}

\begin{figure*}[th]
  \centering
 \includegraphics[width=0.495\textwidth]{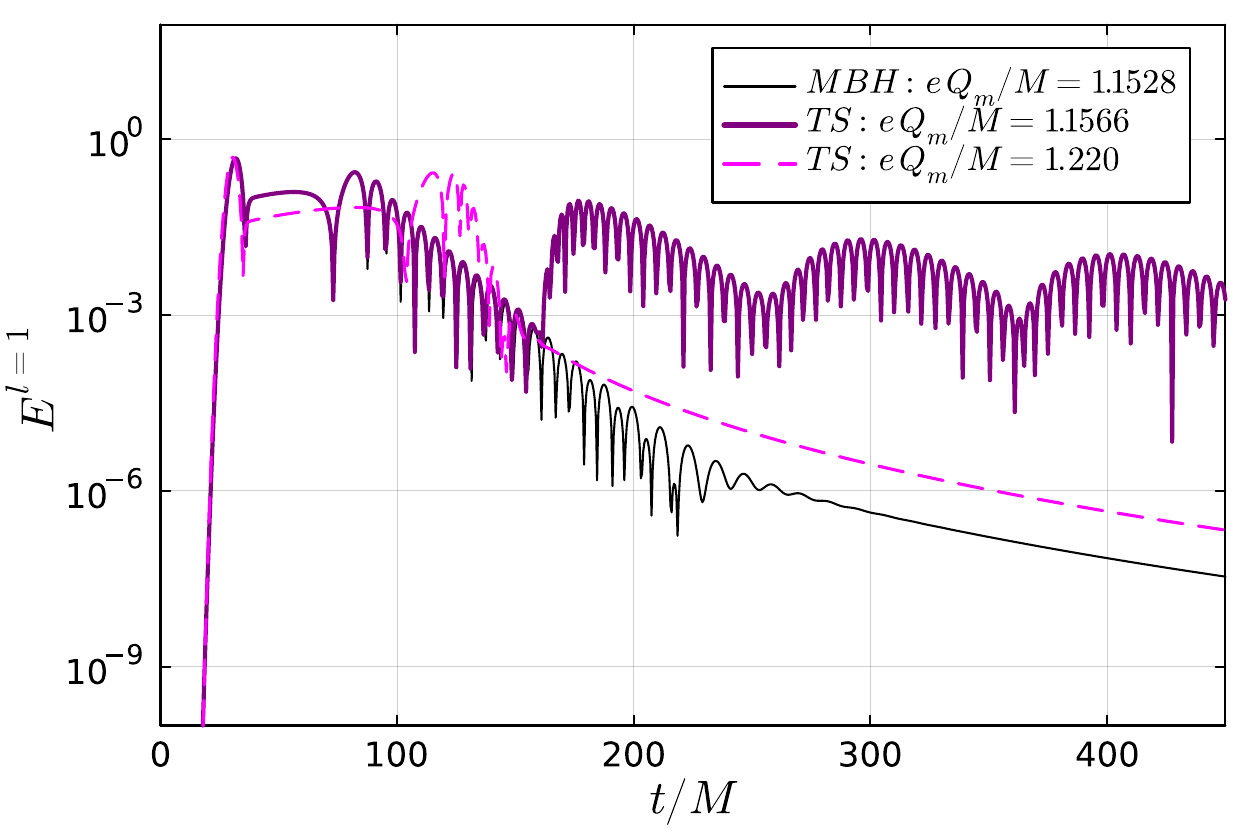}
  \includegraphics[width=0.495\textwidth]{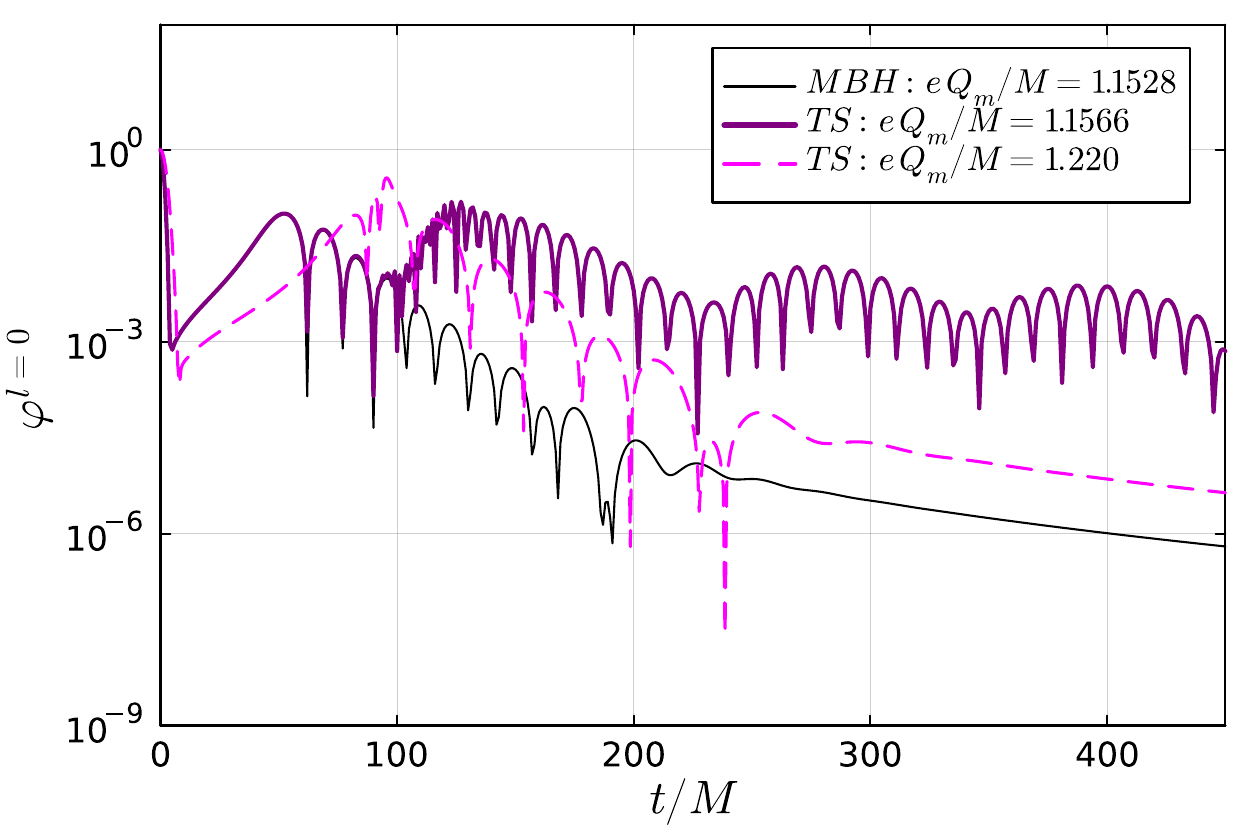}
  \caption{Same as Fig.~\ref{fig:Z12_RingdownVSEchoes} for Type-I $l=1$ perturbations (left panel) and Type-II $l=0$ perturbations (right panel).} \label{fig:E1_RingdownVSEchoes}
\end{figure*}

We shall only show selected cases, since the qualitative features are similar in all sectors. One of our main result was anticipated in Fig.~\ref{fig:Z12_RingdownVSEchoes} for $l=2$ Type-I perturbations, namely odd-parity gravitational and even-parity EM perturbations. In this example we focus on a nearly-extremal magnetized BH with a charge-to-mass ratio similar to that of a second-kind TS. In practice, we consider a BH and a TS solution slightly below and above the $r_B/r_S=1$ threshold, respectively. Results are normalized by the mass of the solution so, in practice, the magnetized BH and the TS have the same mass and a very similar charge (as shown in the phase diagram~\ref{fig:paramspace}, it is not possible to have TSs and BHs with exactly the same charge-to-mass ratio).
In this condition the effective potentials for perturbations of BHs and TSs are very similar at large distances and they remain so even at smaller distances down to the inner region, as shown in Fig.~\ref{fig:potentialBHTS} for the simpler case of $l=1$ Type-I perturbations (similar results apply to other sectors).
In this example the potentials are very similar around the maximum, the shape of which is responsible for the prompt ringdown in the time domain~\cite{Cardoso:2016oxy,Cardoso:2016rao,Cardoso:2017cqb}. However, near the inner boundary the behavior is completely different: the potential vanishes as $r\to r_S$ for the BH while it diverges to positive values as $r\to r_B$ for the TS. The latter behavior supports the long-lived modes discussed in the previous section, which dominate the signal at late times.
This discussion is perfectly consistent with what shown in Fig.~\ref{fig:Z12_RingdownVSEchoes}: the initial ringdown is almost indistinguishable between the BH and TS cases, the small differences are only due to the slightly different charge-to-mass ratio.
However, after the perturbation had  time to probe the inner boundary of the TS and gets reflected, the signal is dominated by echoes associated with perturbations being reflected back and forth between the inner boundary and the unstable photon sphere.
This behavior is generic as long as the charge-to-mass ratio is similar, as shown in Fig.~\ref{fig:E1_RingdownVSEchoes} for the case  of Type-I, $l=1$ perturbations and for $l=0$ perturbations. In the same plot, we also show an example of first-kind TS. Due to the absence of unstable photon sphere, in this case the response does not show long-lived modes and the prompt ringdown is completely different from the BH case, even though the charge-to-mass ratio is only $\approx6\%$ different.

Thus, TSs provide a concrete model, arising as a solution to a consistent theory, in which the time signal smoothly interpolates between different regimes, including one
in which a clean echo signal appears in the gravitational waves. This improves on previous studies of ultracompact objects, which either considered phenomenological backgrounds or test fields, due to the complexity of the field equations (see~\cite{Cardoso:2019rvt} for an overview).

\begin{figure*}[th]
  \centering
  \includegraphics[width=0.495\textwidth]{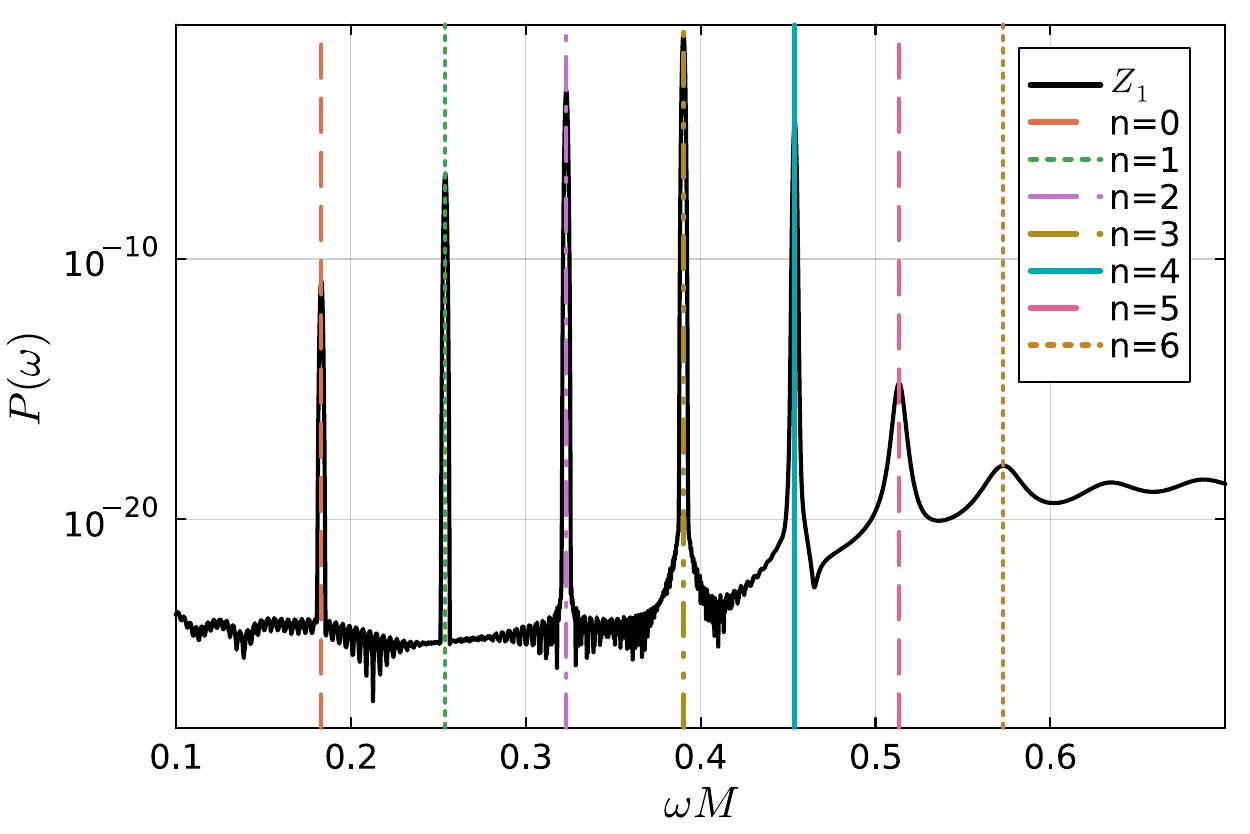}
  \includegraphics[width=0.495\textwidth]{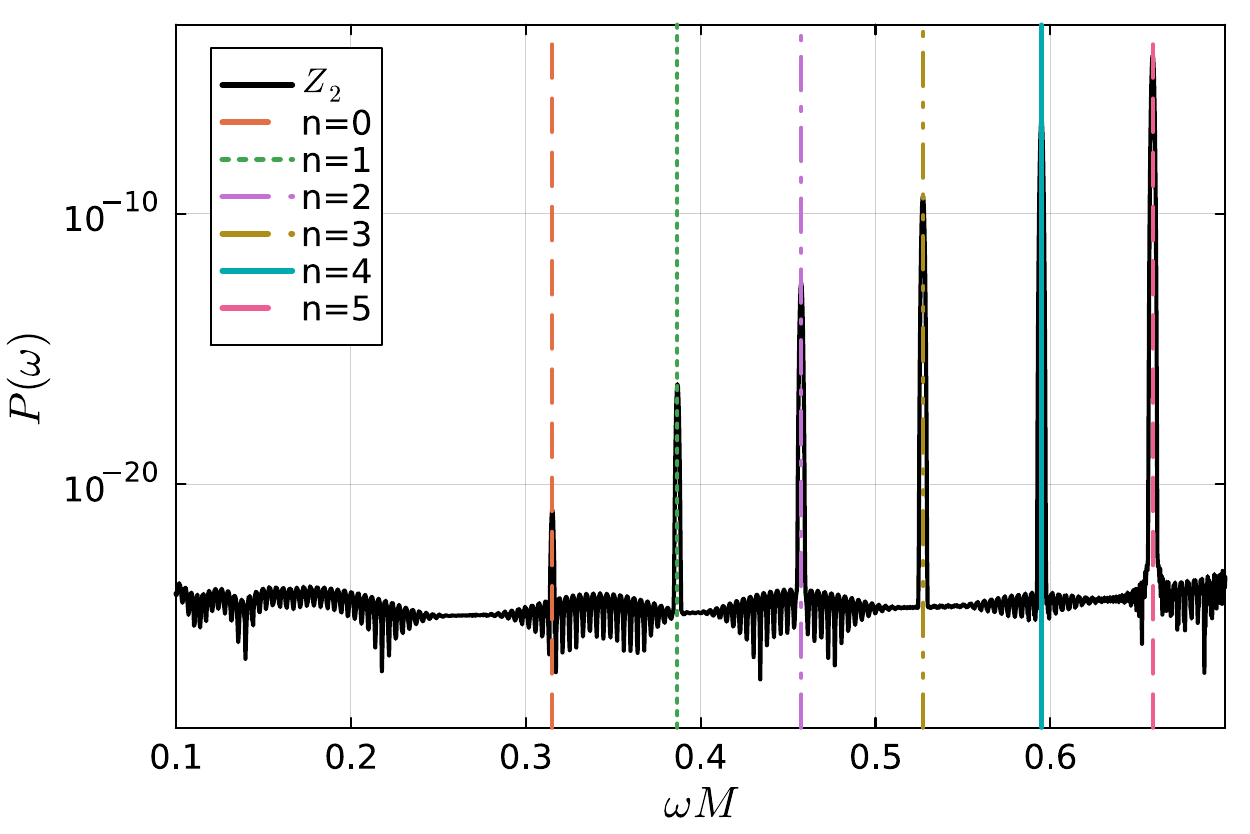}
  \caption{Power spectrum of Type-I perturbations with $l=2$ on a second-kind TS background same charge-to-mass ratio as in Fig.~\ref{fig:Z12_RingdownVSEchoes}. The left and right panel refer to gravitational-driven and EM-driven perturbations, respectively. We highlight the peaks corresponding to the fundamental QNM along with higher overtones up to $n=6$ ($n=5$) of the $Z_1$ ($Z_2$) perturbations.} \label{fig:PWS_Z1_Z2}
\end{figure*}

Finally, in Fig.~\ref{fig:PWS_Z1_Z2} we show an example of power spectrum obtained from the time evolution of Type-I perturbations with $l=2$ on a second-kind TS background. In this case, multiple peaks are present and these allow for a precise estimate of the QNM frequency and damping time for several overtones, as shown in the previous tables.

\section{Conclusions} \label{sec:conclusion}

We thoroughly examined the coupled scalar-EM-gravitational perturbations of magnetized BHs and TSs originating from the dimensional compactification of Einstein-Maxwell theory in five dimensions. 
Our results, supported by both frequency-domain and time-domain analysis, provide strong numerical evidence for the linear stability of these solutions against radial perturbations and axial gravitational perturbations (which are coupled to polar EM ones), in a region of parameter space not affected by the inherited Gregory-Laflamme instability of the black string. In particular, we showed that topological stars with $1<r_B/r_S<2$ are stable for such modes with no dependence on the fifth dimension.
The numerical analysis of the more involved polar gravitational perturbations (which are coupled to axial EM and scalar ones) for $l\geq1$, as well as the case of perturbations with Kaluza-Klein momentum, will appear in a companion paper~\cite{companion}.

Overall, the perturbations that we studied in this paper can all be reduced to a single second-order differential equation and  display qualitatively similar properties. In particular, we confirm the expectation that ultracompact TSs with a stable photon sphere support long-lived modes, giving rise to echoes in the (scalar, EM, and gravitational-wave) signal at late times.
Thus, TSs provide a concrete model, arising as a solution to a consistent theory, in which a clean echo signal appears in the gravitational waves. To the best of our knowledge, this is the first example of a consistent solution\footnote{In this respect it is interesting to remark that, restricting to the four-dimensional perspective, TSs are singular solutions. It is only when considering them as solutions to a higher-dimensional theory that their absence of pathologies becomes evident.} showing echoes in the gravitational-wave signal, since previous studies of ultracompact objects considered either phenomenological backgrounds or test fields, due to the complexity of the field equations.

Although we found no evidence of linear instabilities (at least in the sectors presented in this work and besides the well-known Gregory-Laflamme instability existing only in a certain region of the parameter space), there are arguments suggesting that ultracompact objects might be unstable at the nonlinear level~\cite{Cardoso:2014sna,Cunha:2017qtt}.
This is due to the slow (possibly logarithmic) decay in time, as discussed for microstate geometries~\cite{Eperon:2016cdd} and for other ultracompact objects~\cite{Cardoso:2014sna,Keir:2014oka}.
TSs provide a well-defined model in which the nonlinear evolution of the perturbations can be possibly studied in a relatively simple setting.

We have derived the full set of equations describing the linear response of magnetized BHs and TSs. Besides analyzing in detail the Type~II sector and perturbations with Kaluza-Klein momentum~\cite{companion}, a natural follow-up of our analysis is to study tidal perturbations of these solutions and compute their tidal Love numbers, extending the test scalar case studied in~\cite{Bianchi:2023sfs}. We expect that the various Love numbers of a TS that can be defined in the different perturbation sectors are generically nonzero, and tend to their corresponding value in the extremal BH case as $r_B/r_S\to1$, as it occurs in other models~\cite{Pani:2015tga,Cardoso:2017cfl}.
Another relevant extension is to consider spinning TSs or other topological solitons with less symmetry~\cite{Bah:2022yji}.

\begin{acknowledgments}
We thank Iosif Bena, Massimo Bianchi, Roberto Emparan, Pierre Heidmann, and many participants of  \href{https://indico.in2p3.fr/event/30310/}{Black-Hole Microstructure~VI} (Paris Saclay, 10-15 June 2024) for interesting discussions. 
We are grateful to Giorgio Di Russo and Francisco Morales who have shared their preliminary results with us~\cite{Bena:2024hoh}.
This work is partially supported by the MUR PRIN Grant 2020KR4KN2 ``String Theory as a bridge between Gauge Theories and Quantum Gravity'', by the FARE programme (GW-NEXT, CUP:~B84I20000100001), and by the INFN TEONGRAV initiative.
Some numerical computations have been performed at the Vera cluster supported by the Italian Ministry for Research and by Sapienza University of Rome.
\end{acknowledgments}

\appendix

\begin{widetext}

\section{Linear perturbations in Regge-Wheeler-Zerilli gauge} \label{app:RW}

Regge-Wheeler gauge, even perturbations of the metric:

\begin{align}
h_{\mu\nu}^{\rm even}=\sum_{l,m}\left(
\begin{array}{cccc}
  f_Sf_B^{1/2} H_0(t,r) &  H_1(t,r) & 0 & 0 \\
   H_1(t,r) & f_S^{-1}f_B^{-1/2} H_2(t,r) & 0 & 0 \\
   0 & 0 & r^2 f_B^{1/2} K(t,r) & 0 \\
   0 & 0 & 0 & r^2 f_B^{1/2} \sin\theta^2 K(t,r)
\end{array}
\right)  Y_{lm}(\theta,\phi)\,.
\end{align}

The odd sector of the metric reads
\begin{align}
h_{\mu\nu}^{\rm odd}=\sum_{l,m}\left(
\begin{array}{cccc}
  0 & 0 & -h_0(t,r)/\sin\theta \partial_\phi & h_0(t,r) \sin\theta \partial_\theta \\
  0 & 0 & -h_1(t,r)/\sin\theta \partial_\phi & h_1(t,r) \sin\theta \partial_\theta \\
  -h_0(t,r)/\sin\theta \partial_\phi & -h_1(t,r)/\sin\theta \partial_\phi & 0 & 0 \\
   h_0(t,r) \sin\theta \partial_\theta & h_1(t,r) \sin\theta \partial_\theta  & 0 & 0
\end{array}
\right) Y_{lm}(\theta,\phi)
\end{align}

The even-parity EM perturbations (for either $F_{\mu\nu}$ or $\F_{\mu\nu}$) read
\begin{align}~\label{eq:rw_ansatz_even_em}
f_{\mu\nu}^{\rm even} & =\sum_{l,m}\left(
\begin{array}{cccc}
   0 & f_{01}^+(t,r) & f_{02}^+(t,r) \partial_\theta & f_{02}^+(t,r) \partial_\phi \\
  -f_{01}^+(t,r) & 0 & f_{12}^+(t,r) \partial_\theta & f_{12}^+(t,r) \partial_\phi \\
  -f_{02}^+(t,r) \partial_\theta  & -f_{12}^+(t,r) \partial_\theta  & 0 & 0  \\
  -f_{02}^+(t,r) \partial_\phi & -f_{12}^+(t,r) \partial_\phi  & 0 & 0
\end{array}
\right) Y_{lm}(\theta,\phi)
\\
g_{\mu\nu}^{\rm even} & =\sum_{l,m}\left(
\begin{array}{cccc}
   0 & g_{01}^+(t,r) & g_{02}^+(t,r) \partial_\theta & g_{02}^+(t,r) \partial_\phi \\
  -g_{01}^+(t,r) & 0 & g_{12}^+(t,r) \partial_\theta & g_{12}^+(t,r) \partial_\phi \\
  -g_{02}^+(t,r) \partial_\theta  & -g_{12}^+(t,r) \partial_\theta  & 0 & 0  \\
  -g_{02}^+(t,r) \partial_\phi & -g_{12}^+(t,r) \partial_\phi  & 0 & 0
\end{array}
\right) Y_{lm}(\theta,\phi)
\end{align}
while the odd-parity EM perturbations are
\begin{align}~\label{eq:rw_ansatz_odd_em}
f_{\mu\nu}^{\rm odd} & =\sum_{l,m}\left(
\begin{array}{cccc}
  0 & 0 & f_{02}^-(t,r)/\sin\theta \partial_\phi & -f_{02}^-(t,r) \sin\theta \partial_\theta \\
  0 & 0 & f_{12}^-(t,r)/\sin\theta \partial_\phi & -f_{12}^-(t,r) \sin\theta \partial_\theta \\
  -f_{02}^-(t,r)/\sin\theta \partial_\phi & -f_{12}^-(t,r)/\sin\theta \partial_\phi & 0 & f_{23}^-(t,r) \sin\theta  \\
   f_{02}^-(t,r) \sin\theta \partial_\theta & f_{12}^-(t,r) \sin\theta \partial_\theta  & -f_{23}^-(t,r) \sin\theta & 0
\end{array}
\right) Y_{lm}(\theta,\phi)
\\
g_{\mu\nu}^{\rm odd} & =\sum_{l,m}\left(
\begin{array}{cccc}
  0 & 0 & g_{02}^-(t,r)/\sin\theta \partial_\phi & -g_{02}^-(t,r) \sin\theta \partial_\theta \\
  0 & 0 & g_{12}^-(t,r)/\sin\theta \partial_\phi & -g_{12}^-(t,r) \sin\theta \partial_\theta \\
  -g_{02}^-(t,r)/\sin\theta \partial_\phi & -g_{12}^-(t,r)/\sin\theta \partial_\phi & 0 & g_{23}^-(t,r) \sin\theta  \\
   g_{02}^-(t,r) \sin\theta \partial_\theta & g_{12}^-(t,r) \sin\theta \partial_\theta  & -g_{23}^-(t,r) \sin\theta & 0
\end{array}
\right) Y_{lm}(\theta,\phi)
\end{align}

Finally, scalar perturbations are simply decomposed as
\begin{align}~\label{eq:rw_ansatz_scal}
  \delta\Phi & = \sum_{l,m}\frac{\varphi(t,r)}{r}  Y_{lm}(\theta,\phi)
  \\
  \delta\Xi & = \sum_{l,m}\frac{\xi(t,r)}{r}  Y_{lm}(\theta,\phi)
\end{align}
Note that, for clarity, we have omitted the indices $(l,m)$ in the coefficients (which are functions of $t$ and $r$) of the spherical-harmonic decomposition. The symmetry of the background guarantees that $m$ is degenerate and perturbations with different values of $l$ are decoupled from each other.

\section{Type-II equations}\label{app:type_II}

\begin{align}
    E_{tt} & = 
      f_Bf_S^2 \partial_r^2 K
    + \left(\frac{3f_Bf_S^2}{r} + f_S^2 f_B' + \frac{1}{2}f_Bf_Sf_S'\right) \partial_r K
    - \frac{f_S^2}{4r}\left( 4f_B+rf_B'\right)\partial_rH_2
    + \frac{\sqrt{3}}{4r}f_S^2f_B'\partial_r\varphi
    \nonumber\\
    &
    - \left( \frac{\Lambda f_S}{2r^2} + \frac{f_Bf_S^2}{r^2} + \frac{f_S^2f_B'}{r} + \frac{f_Sf_S'}{4r}( 4 f_B + rf_B')\right) H_2
    + \left(\frac{f_S}{r^2} - \frac{Q_m^2\kappa_4^2f_S}{r^4} - \frac{\Lambda f_S}{2r^2}\right) K
    \nonumber\\
    &
    - \left( \frac{\sqrt{3}f_S^2f_B'}{4r^2}-\frac{\sqrt{3}Q_m^2\kappa_4^2f_S}{6r^5}\right) \varphi
    + \frac{4Q_m\kappa_4^2f_S}{er^3} f_{23}^- 
    = 0
    \label{eq:Ett}
    \\
    E_{tr} & =  
      4 r f_B \partial_r K
    + \frac{2f_B}{f_S}(2f_S-rf_S') K
    + \frac{2\Lambda \sqrt{f_B}}{\I \omega r} H_1
    + \sqrt{3}f_B'\varphi 
    - (4f_B+rf_B')H_2
    = 0
    \label{eq:Etr}
    \\
    E_{t\theta} & = 
      \frac{f_S\sqrt{f_B}}{\I \omega} \partial_r H_1
    + \frac{\left( f_Sf_B' + 2 f_B f_S' \right) }{2 \I \omega \sqrt{f_B}} H_1
    + H_2 + K
    - \frac{2Q_m\kappa_4^2}{e r^2 \Lambda} f_{23}^- 
    = 0
    \label{eq:Ettheta}
    \\
    E_{rr} & = 
      f_S^2\left(4f_B+rf_B'\right) \partial_r H_0
    - 2f_S\left(2f_Bf_S+rf_Sf_B'+rf_Bf_S'\right)\partial_r K
    + \sqrt{3}f_S^2f_B' \partial_r\varphi
    \nonumber\\
    &
    - \frac{2\Lambda f_S}{r} H_0
    + f_S\left( \frac{4f_Bf_S}{r} + 4f_Sf_B' + 4f_Bf_S' + r f_B'f_S' \right) H_2
    + 2\I\omega\frac{f_S}{\sqrt{f_B}} \left(4f_B+r f_B'\right) H_1
    \nonumber\\
    &
    + 2r\left( \frac{2 Q_m^2\kappa_4^2 f_S}{r^4} + \frac{(\Lambda-2)f_S}{r^2} - 2\omega^2\right) K
    - \left( \frac{\sqrt{3}f_S^2f_B'}{r} + \frac{2\sqrt{3}Q_m^2\kappa_4^2f_S}{3r^4}\right) \varphi
    - \frac{4Q_m\kappa_4^2f_S}{er^3} f_{23}^- 
    = 0
    \label{eq:Err}
    \\
    E_{r\theta} & = 
      f_Bf_S \left(\partial_r H_0 - \partial_r K \right)
    + \frac{2Q_m\kappa_4^2}{e\Lambda} \frac{f_Bf_S}{r^2} \partial_rf_{23}^-
    - \frac{1}{2}\left(\frac{2f_Bf_S}{r} - f_Bf_S'\right) H_0
    + \frac{1}{2}\left(\frac{2f_Bf_S}{r} + f_Sf_B' + f_Bf_S' \right) H_2
    \nonumber\\
    &
    + \I \omega \sqrt{f_B} H_1
    - \frac{\sqrt{3}f_Sf_B'}{2r} \varphi 
    =0
    \label{eq:Ertheta}
    \\
    E_{\theta\theta} & = 
      f_Bf_S^2 \left(\partial_r^2 H_0 - \partial_r^2 K \right)
    + \left( \frac{f_Bf_S^2}{r} + f_S^2f_B' + \frac{3}{2}f_Bf_Sf_S' \right) \partial_r H_0
    + 2\I \omega \sqrt{f_B} f_S \partial_r H_1
    \nonumber\\
    &
    + \left( \frac{f_Bf_S^2}{r} + \frac{1}{2}f_S^2f_B' + \frac{1}{2}f_Bf_Sf_S' \right) \partial_r H_2
    - \left( \frac{2f_Bf_S^2}{r} + f_S^2f_B' + f_Bf_Sf_S' \right) \partial_r K
    - \frac{\sqrt{3}f_S^2f_B'}{2r} \partial_r\varphi
    \nonumber\\
    &
    + \frac{\I \omega }{\sqrt{f_B}} \left( \frac{2f_Bf_S}{r} + f_Sf_B' + f_Bf_S'\right) H_1
    - \left(\omega^2 - \frac{3}{2}f_Sf_B'f_S'\right) H_2
    - \left(\omega^2 + \frac{2Q_m^2\kappa_4^2f_S}{r^4} \right) K
    \nonumber\\
    &
    + \left( \frac{\sqrt{3}f_S^2f_B'}{2r^2} + \frac{\sqrt{3}Q_m^2\kappa_4^2f_S}{3r^5}\right)\varphi
    + \frac{2Q_m\kappa_4^2f_S}{e r^4}f_{23}^-
    =0
    \label{eq:Ethetatheta}
    \\
    E_{\theta\phi} & = 
    H_0 - H_2  = 0
    \label{eq:Ethetaphi}
    \\
    E_{\varphi} & =
     f_Bf_S^2\partial_r^2\varphi
    + \left(f_S^2f_B' + f_Bf_Sf_S'\right) \partial_r \varphi
    + \frac{\sqrt{3}}{2}r f_S^2 f_B' \left( \partial_r K - \partial_r H_2 \right)
    \nonumber\\
    &
    + \left(\omega^2 - \frac{\Lambda f_S}{r^2} -\frac{Q_m^2\kappa_4^2f_S}{3r^4}-\frac{f_S^2f_B'+f_Bf_Sf_S'}{r}\right) \varphi
    - \frac{\sqrt{3} \I \omega r f_Sf_B'}{2\sqrt{f_B}} H_1
    - \frac{\sqrt{3} r f_Sf_B'f_S'}{2} H_2
    \nonumber\\
    &
    + \frac{2\sqrt{3} Q_m^2\kappa_4^2 f_S}{3r^3} K
    - \frac{2\sqrt{3}Q_m\kappa_4^2f_S}{3e r^3} f_{23}^- 
    = 0
    \label{eq:scalar}
    \\
    E_{f23m} & = 
     f_Bf_S^2\partial_r^2f_{23}^-
    + \left(f_S^2f_B'+f_Bf_Sf_S'\right) \partial_r f_{23}^-
    + \left(\omega^2-\frac{\Lambda f_S}{r^2}\right) f_{23}^-
    + \frac{eQ_m\Lambda f_S}{r^2} K 
    - \frac{\sqrt{3} eQ_m\Lambda f_S}{3r^3} \varphi 
    = 0
    \label{eq:em}
\end{align}

\section{Decoupled equations}\label{app:decoupled}

For completeness, in this appendix we provide the field equations for the perturbations of ${\cal F}$ and $\Xi$ on the background of a magnetized BH or TS. Since ${\cal F}=0=\Xi$ on these background, the field equations decouple at the linear level and can be written as those of a test scalar, $\xi := f_B^{-1/4}\delta \Xi$, and a test massless gauge with with two physical degrees of freedom, $\mathfrak{E}:= f_B^{5/4}f_S g_{12}^+$ and $\mathfrak{B}:= f_B^{3/4} g_{23}^-$, respectively. These can be derived by considering Eqs.~\eqref
{eq:EMSxi} and~\eqref{eq:EMScurlyFmunu} and linear perturbations defined in Eqs.~\eqref{eq:rw_ansatz_even_em},~\eqref{eq:rw_ansatz_odd_em} and ~\eqref{eq:rw_ansatz_scal}:

\begin{align}
 & {\cal D}[\xi]  
+ \left(\frac{1}{2}f_S^2f_B' + f_Bf_Sf_S' \right) \partial_r \xi
- \left(
 \frac{\Lambda f_S}{r^2} 
+ \frac{f_Bf_Sf_S'}{r}
+ \frac{f_S^2f_B'}{2r}
+\frac{3 f_S^2f_B'^2}{16 f_B}
- \frac{f_Sf_B'f_S'}{4}
\right)\xi
 = 0 \,,
 \label{eq:decoup_xi}
 \\
 & {\cal D}[\gE]  
+ \left(\frac{1}{2}f_S^2f_B' + f_Bf_Sf_S' \right)\partial_r \gE
- \left(
 \frac{\Lambda f_S}{r^2} 
+ \frac{3f_S^2f_B'}{2r}
+ \frac{15 f_S^2 f_B'^2}{16 f_B}
- \frac{3f_Sf_B'f_S'}{4}
\right)\gE
 = 0 \,,
 \label{eq:decoup_gothE}
 \\
 & {\cal D}[\gB]  
+ \left(\frac{1}{2}f_S^2f_B' + f_Bf_Sf_S' \right)\partial_r \gB
- \left(
 \frac{\Lambda f_S}{r^2} 
- \frac{3f_S^2f_B'}{2r}
+ \frac{3f_S^2f_B'^2}{16 f_B}
+ \frac{3f_Sf_B'f_S'}{4}
\right)\gB
 = 0 \,.
 \label{eq:decoup_gothB}
\end{align}

These equations can be cast in Schrodinger-like form by transforming to the $\rho$ coordinate:
\begin{align}
 & \partial_\rho^2 \xi 
+ \left[ \omega^2 
- f_S\left(
 \frac{\Lambda }{r^2} 
+ \frac{f_Bf_S'}{r}
+ \frac{f_Sf_B'}{2r}
+\frac{3 f_Sf_B'^2}{16 f_B}
- \frac{f_B'f_S'}{4}
\right)\right]\xi
 = 0 \,,
 \label{eq:decoup_xib}
 \\
 &  \partial_\rho^2 \gE
+ \left[ \omega^2 
- f_S\left( \frac{\Lambda }{r^2} 
+ \frac{3f_S f_B'}{2r}
+ \frac{15 f_S f_B'^2}{16 f_B}
- \frac{3f_B'f_S'}{4}
\right)\right]\gE
 = 0 \,,
 \label{eq:decoup_gothEb}
 \\
 & \partial_\rho^2 \gB 
+ \left[\omega^2
- f_S\left(
 \frac{\Lambda }{r^2} 
- \frac{3f_Sf_B'}{2r}
+ \frac{3f_Sf_B'^2}{16 f_B}
+ \frac{3f_B'f_S'}{4}
\right)\right]\gB
 = 0 \,.
 \label{eq:decoup_gothBb}
\end{align}

All three effective potentials above are singular at $r=r_B$. From Eqs.~\eqref{eq:decoup_xi}--\eqref{eq:decoup_gothB} the following indicial equations can be derived:
\begin{align}
  16\lambda_\xi^2 -8 \lambda_\xi -3
 = 0\,,\qquad 16\lambda_\gE^2 -8 \lambda_\gE -15
 = 0\qquad  16\lambda_\gB^2 -8 \lambda_\gB -3
 = 0\,.
 \label{eq:indicial_dec}
\end{align}

From the latter, one can deduce that the field equations admit regular solutions corresponding to $\lambda_\xi=\frac{3}{4}=\lambda_\gB$, and $\lambda_\gE=\frac{5}{4}$.

\section{Convergence tests of time-domain codes} \label{app:convergence}

To check the convergence with resolution of our results, we have simulated the time evolution of $\lbrace Z_1, Z_2 \rbrace$ perturbations (see Sec.~\ref{sec:eqTypeI}) of a second-kind TS ($r_B =1.01 \,r_S$) with low (L), medium (M) and high (H) resolution, corresponding to $N=\lbrace 2^{16}, 2^{17}, 2^{18}\rbrace$ points on the radial grid.

In Fig.~\ref{fig:conv_plots} we plot the relative error between medium and low resolution results, $L-M$, and between high and medium resolution time series, $Q_n (M-H)$, rescaled by the appropriate convergence factor:

\begin{align}
    Q_n := \frac{(dr_{L})^n-(dr_M)^n}{(dr_M)^n-(dr_H)^n}
\end{align}

From the comparison in Fig.~\ref{fig:conv_plots} we notice that during the prompt ringdown phase the convergence matches the expectations (i.e. is compatible with fourth order convergence), but undergoes a sudden drop in the convergence order during the first reflection of the signal on the surface of TS. In the subsequent phase of the signal, the convergence order is then restored.

One possible culprit for the loss in convergence during the first reflection could consist in the finite difference approximation of the regularity BC we impose at the TS boundary, although the latter is realized with a fourth-order stencil. Alternatively, we suspect that resolving the regular singular point at $r=r_B$ requires much higher resolution, while our current results are only marginally in the convergence regime. 

On the other hand, this underperformance of our time-domain framework in resolving the boundary of the star does not seem to affect the robustness of the results since the afflicted part of the signal contributes only marginally to the full spectrum.
Indeed, most of the information we extract with our spectral analysis is enclosed in the part of the signal that follows the first reflection of the initial wave packet. Here the tortoise coordinate we choose allows resolving sufficiently well the effective potential and its cavity, which explains the restoration in the convergence order we observe.

\begin{figure}[th]
  \centering
  \includegraphics[width=0.495\textwidth]{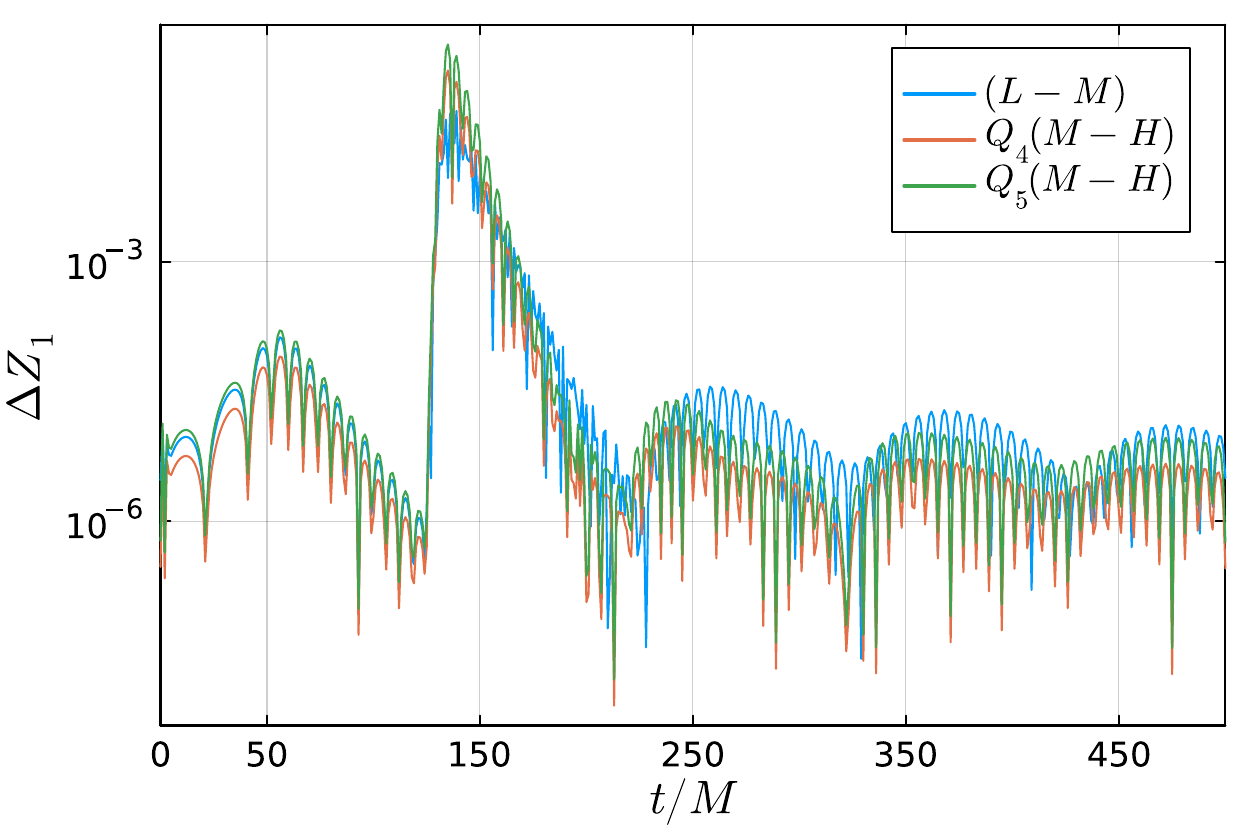}
    \includegraphics[width=0.495\textwidth]{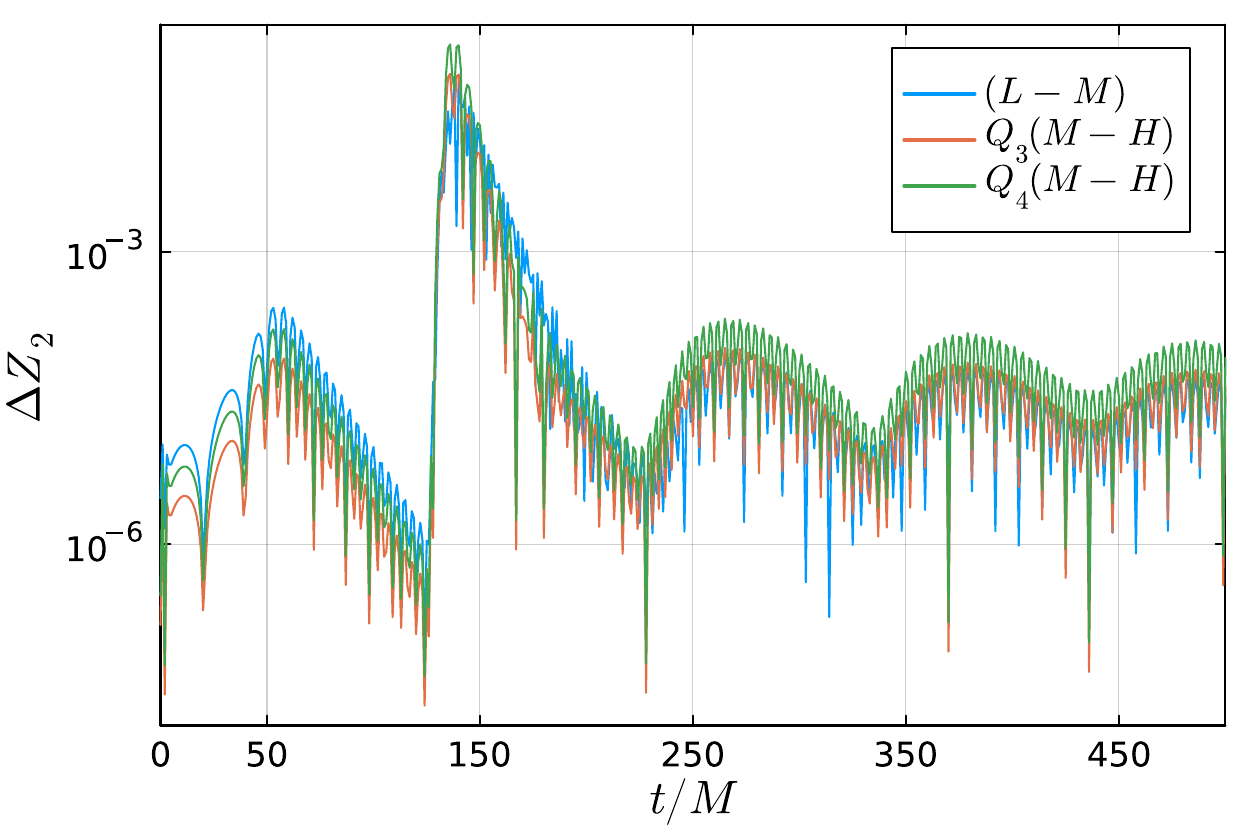}
  \caption{Convergence test of our $1+1$ integrator, see text for details.  
 } \label{fig:conv_plots}
\end{figure}

\end{widetext}

\clearpage
\bibliographystyle{apsrev4-1}
\bibliography{mbh_topstar_spectroscopy.bib}

\end{document}